\documentclass[longauth]{aa}  

\usepackage{graphicx}
\usepackage{siunitx}
\usepackage{natbib}
\usepackage{amsmath}
\usepackage{tabularx}
\usepackage{float}
\usepackage{placeins}
\usepackage{fancyvrb}
\usepackage{txfonts}
\usepackage[dvipsnames]{xcolor}
\usepackage[breaklinks, colorlinks, citecolor=Cerulean]{hyperref}
\usepackage{orcidlink}

\newcommand{\Hb}{\hbox{H$\beta$}}
\newcommand{\Ha}{\hbox{H$\alpha$}}
\newcommand{\oi}{[O~\textsc{i}]$\lambda6300$}
\newcommand{\oiii}{[O~\textsc{iii}]$\lambda5007$}
\newcommand{\nii}{[N~\textsc{ii}]$\lambda6584$}

\newcommand{\sii}{[S~\textsc{ii}]$\lambda\lambda6717,6731$}

\newcommand{\hei}{He~\textsc{i}~$\lambda5875$}
\newcommand{\heii}{He~\textsc{ii}~$\lambda4685$}

\newcommand{\oion}{[O~\textsc{i}]}

\newcommand{\oiiion}{[O~\textsc{iii}]}
\newcommand{\niion}{[N~\textsc{ii}]}
\newcommand{\siiion}{[S~\textsc{iii}]}
\newcommand{\siion}{[S~\textsc{ii}]}
\newcommand{\heion}{He~\textsc{i}}
\newcommand{\heiion}{He~\textsc{ii}}

\newcommand{\ngc}{NGC~}
\newcommand{\Msun}{$\mbox{M}_{\sun}$}
\newcommand{\Lsun}{$\mbox{L}_{\sun}$}
\newcommand{\hii}{H~{\sc ii}}
\newcommand{\sn}{S$/$N}
\newcommand{\ebv}{$E(B-V)$}

\DeclareSIUnit\angstrom{\text {\AA}}
\DeclareSIUnit\parsec{pc}
\DeclareSIUnit\lightyear{ly}
\DeclareSIUnit\year{yr}
\DeclareSIUnit\mag{mag}
\DeclareSIUnit\erg{erg}
\DeclareSIUnit\Msun{M_\odot}
\DeclareSIUnit\Lsun{L_\odot}
\DeclareSIUnit\Zsun{Z_\odot}
\DeclareSIUnit\dex{dex}
\DeclareSIUnit\percent{per cent}
\DeclareSIUnit\arcmin{arcmin}
\DeclareSIUnit\arcsec{arcsec}
\DeclareSIUnit\pix{px}

\begin{document} 
\VerbatimFootnotes

   \title{The MUSE view of the Sculptor galaxy: survey overview and the planetary nebulae luminosity function.}
    \titlerunning{NGC 253 PNLF}

\author{E. Congiu\thanks{econgiu@eso.org}\orcidlink{0000-0002-8549-4083}\inst{\ref{ESO}},
F. Scheuermann\orcidlink{0000-0003-2707-4678}\inst{\ref{ARI}},
K. Kreckel\inst{\ref{ARI}},
A. Leroy\inst{\ref{OSU},\ref{CCAP}},
E. Emsellem\orcidlink{0000-0002-6155-7166}\inst{\ref{ESO2}, \ref{lyon}},
F. Belfiore\orcidlink{0000-0002-2545-5752}\inst{\ref{INAF}},
J. Hartke\orcidlink{0000-0002-8745-689X}\inst{\ref{FINCA}, \ref{Tuorla}, \ref{TCSMT}},
G. Anand\orcidlink{0000-0002-5259-2314}\inst{\ref{STSCI}},
O.~V.~Egorov\orcidlink{0000-0002-4755-118X}\inst{\ref{ARI}},
B. Groves\inst{\ref{ICRA}},
T. Kravtsov\inst{\ref{Tuorla},\ref{FINCA}},
D. Thilker\inst{\ref{Hopkins}},
C. Tovo\inst{\ref{UNIPD}},
F. Bigiel\orcidlink{0000-0003-0166-9745}\inst{\ref{AIfA}},
G. A. Blanc\orcidlink{0000-0003-4218-3944}\inst{\ref{Carnegie},\ref{UChile}},
A. D. Bolatto\orcidlink{0000-0002-5480-5686}\inst{\ref{UMD}},
S. A. Cronin\orcidlink{0000-0002-9511-1330}\inst{\ref{UMD}},
D. A. Dale\orcidlink{0000-0002-5782-9093}\inst{\ref{Uwyoming}},
R. McClain\orcidlink{0000-0002-6187-4866}\inst{\ref{OSU},\ref{CCAP}},
J. E. M\'endez-Delgado\orcidlink{0000-0002-6972-6411}\inst{\ref{UNAM}},
E. K. Oakes\orcidlink{0000-0002-0119-1115}\inst{\ref{UConn}},
R. S.\ Klessen\orcidlink{0000-0002-0560-3172}\inst{\ref{ZAH},\ref{IWR},\ref{Harvard},\ref{HRI}},
E. Schinnerer\orcidlink{0000-0002-3933-7677}\inst{\ref{MPIA}},
T. G. Williams\orcidlink{0000-0002-0012-2142}\inst{\ref{Oxford}}.
}

\authorrunning{Congiu et al.}

\institute{
European Southern Observatory (ESO), Alonso de Córdova 3107, Casilla 19, Santiago 19001, Chile; \label{ESO} \and
Astronomisches Rechen-Institut, Zentrum f\"{u}r Astronomie der Universit\"{a}t Heidelberg, M\"{o}nchhofstra\ss e 12-14, 69120 Heidelberg, Germany; \label{ARI}\and
Department of Astronomy, The Ohio State University, 140 West 18th Avenue, Columbus, OH 43210, USA; \label{OSU}\and
Center for Cosmology and Astroparticle Physics, 191 West Woodruff Avenue, Columbus, OH 43210, USA; \label{CCAP}\and
European Southern Observatory, Karl-Schwarzschild Stra{\ss}e 2, D-85748 Garching bei M\"{u}nchen, Germany; \label{ESO2} \and
Univ Lyon, Univ Lyon1, ENS de Lyon, CNRS, Centre de Recherche Astrophysique de Lyon UMR5574, F-69230 Saint-Genis-Laval France;\label{lyon} \and
INAF -- Osservatorio Astrofisico di Arcetri, Largo E. Fermi 5, I-50157 Firenze, Italy; \label{INAF} \and
Finnish Centre for Astronomy with ESO, (FINCA), University of Turku, FI-20014 Turku, Finland;\label{FINCA} \and
Tuorla Observatory, Department of Physics and Astronomy, University of Turku, FI-20014 Turku, Finland;\label{Tuorla}\and
Turku Collegium for Science, Medicine and Technology, University of Turku, FI-20014 Turku, Finland;\label{TCSMT}\and
Space Telescope Science Institute, 3700 San Martin Drive, Baltimore, MD 21218, USA;\label{STSCI}\and
International Centre for Radio Astronomy Research, University of Western Australia, 7 Fairway, Crawley, 6009 WA, Australia; \label{ICRA}\and
Department of Physics and Astronomy, The Johns Hopkins University, Baltimore, MD 21218, USA; \label{Hopkins}\and
Dipartimento di Fisica e Astronomia ‘G. Galilei’, Universit`a di Padova, Vicolo dell’Osservatorio 3, I-35122 Padova, Italy;\label{UNIPD} \and
Argelander-Institut f\"ur Astronomie, Universit\"at Bonn, Auf dem H\"ugel 71, 53121 Bonn, Germany; \label{AIfA}\and
Observatories of the Carnegie Institution for Science, 813 Santa Barbara Street, Pasadena, CA 91101, USA; \label{Carnegie}\and 
Departamento de Astronom\'{i}a, Universidad de Chile, Camino del Observatorio 1515, Las Condes, Santiago, Chile; \label{UChile} \and
Department of Astronomy, University of Maryland, College Park, MD 20742, U.S.A.; \label{UMD} \and
Department of Physics and Astronomy, University of Wyoming, Laramie, WY 82071, USA; \label{Uwyoming}\and
Instituto de Astronom\'{\i}a, Universidad Nacional Aut\'onoma de M\'exico, Ap. 70-264, 04510 CDMX, Mexico; \label{UNAM}\and
Department of Physics, University of Connecticut, 196A Auditorium Road, Storrs, CT 06269, USA; \label{UConn}\and
Universit\"{a}t Heidelberg, Zentrum f\"{u}r Astronomie, Institut f\"{u}r Theoretische Astrophysik, Albert-Ueberle-Str.\ 2, 69120 Heidelberg, Germany; \label{ZAH}\and
Universit\"{a}t Heidelberg, Interdisziplin\"{a}res Zentrum f\"{u}r Wissenschaftliches Rechnen, Im Neuenheimer Feld 225, 69120 Heidelberg, Germany; \label{IWR}\and
Harvard-Smithsonian Center for Astrophysics, 60 Garden Street, Cambridge, MA 02138, U.S.A.; \label{Harvard}\and
Elizabeth S. and Richard M. Cashin Fellow at the Radcliffe Institute for Advanced Studies at Harvard University, 10 Garden Street, Cambridge, MA 02138, U.S.A.; \label{HRI}\and
Max-Planck-Institut f\"{u}r Astronomie, K\"{o}nigstuhl 17, D-69117, Heidelberg, Germany; \label{MPIA}\and
Sub-department of Astrophysics, Department of Physics, University of Oxford, Keble Road, Oxford OX1 3RH, UK; \label{Oxford}.
}

   \date{}
 
  \abstract{\ngc253, the Sculptor galaxy, is the southern, massive, star-forming disk galaxy closest to the Milky Way. 
  In this work, we present a new 103-pointing MUSE mosaic of this galaxy covering the majority of its star-forming disk up to $0.75\times$R$_{25}$.
  With an area of $\sim20\times5\,\si{arcmin^2}$ ($\sim20\times5\,\si{kpc^2}$, projected) and a physical resolution of $\sim 15\,\si{pc}$, this mosaic constitutes one of the largest, highest physical resolution integral field spectroscopy surveys of any star-forming galaxy to date. 
  Here, we exploit the mosaic to identify a sample of $\sim$ 500 planetary nebulae ($\sim 20$ times larger than in previous studies) to build the planetary nebula luminosity function (PNLF) and obtain a new estimate of the distance to \ngc253.
  The value obtained is $17\%$ higher than estimates returned by other reliable measurements, mainly obtained via the top of the red giant branch method (TRGB).
  The PNLF also varies between the centre (r $< 4~\si{kpc}$) and the disk of the galaxy.
  The distance derived from the PNLF of the outer disk is comparable to that of the full sample, while the PNLF of the centre returns a distance $\sim$ \SI{0.9}{Mpc} larger.
  Our analysis suggests that extinction related to the dust-rich interstellar medium and edge-on view of the galaxy (the average \ebv\ across the disk is $\sim 0.35~\si{mag}$) plays a major role in explaining both the larger distance recovered from the full PNLF and the difference between the PNLFs in the centre and in the disk.
  }


   \maketitle
%
\section{Introduction}

\ngc253, also known as the Sculptor galaxy, is one of the closest \citep[D$\sim\SI{3.5}{Mpc}$;][]{Newman24, Okamoto24} massive \citep[M$_{star}\sim 4.4 \times 10^{10} \si{M_{\odot}}$;][]{Bailin11, Leroy21} star-forming galaxies to the Milky Way.
It is also one of the largest galaxies in the sky, with an apparent size of 42$\times$\SI{12}{\arcmin^2} \citep{Jarrett19}.
With its prominent stellar bar, well-defined spiral arms \citep{Iodice14}, and star formation spread across the disk, NGC~253 represents a nearby, archetypal example of a main-sequence spiral galaxy.

Its star formation rate (SFR) has been estimated to be between $\sim$4.9 \citep{Leroy19} and $\sim$ \SI{6.5}{M_{\odot}.yr^{-1}} \citep{Jarrett19}.
Almost a third of it \citep[$\sim \SI{2}{M_{\odot}.yr^{-1}}$;][]{Bendo15, Leroy15} is produced by 10 giant molecular clouds distributed in a starburst ring \citep[$\sim 500~ \si{pc}$ in diameter;][]{Leroy15} around its nucleus, making this object the closest massive starburst to our Galaxy.
In addition, the starburst is responsible for launching a powerful multi-phase outflow \citep[e.g.][]{Strickland00,Bauer07,Westmoquette11, Bolatto13, Walter17, Lopez23, Cronin25} that is ejecting large amounts of gas \citep[14--$\SI{39}{M_{\odot}.yr^{-1}}$ for the molecular component only,][]{Krieger19} from the centre of the galaxy into its circumgalactic medium.
Part of this gas seems to remain in the gravitational well of the galaxy, where it cools down to the neutral phase and moves towards the outskirts of the disk \citep{Lucero15}, potentially reaccreting and fueling future star formation.

The proximity of the galaxy allows for excellent spatial resolution to be achieved even with ground-based telescopes given the $17\,\si{pc.arcsec^{-1}}$ scale.
However, when combined with the large intrinsic size of the galaxy (see Table ~\ref{tab:deproj}), this proximity also results in a large apparent size, causing studies in the last decades to focus mostly on the smaller central structures, such as the starburst \citep[e.g.][]{Bendo15, Leroy15} and its related outflow \citep[e.g.][]{Strickland00,Bauer07,Westmoquette11, Bolatto13, Walter17, Lopez23}.

In this galaxy, we have the possibility to resolve the individual components of the star formation process (giant molecular clouds, \hii\ regions, their ionising sources) for thousands of sources distributed across a wide variety of environments (centre, bar, outflow, spiral arms, inter-arm regions) with a resolution that is a factor of 2--5 better than for most of the latest integral field unit (IFU) surveys of nearby galaxies (e.g., SIGNALS: \citealt{Rousseau18}; MAD: \citealt{Erroz-Ferrer19}; TYPHOON: \citealt{Grasha22}; PHANGS-MUSE: \citealt{Emsellem22}), and 50--100 times better than the previous generation large IFU surveys (e.g., CALIFA: \citealt{Sanchez12, Husemann13}; MaNGA: \citealt{Wake17, Bundy15}; SAMI: \citealt{Croom12}).

Large IFU mosaics of extragalactic objects with such high resolution have been obtained so far only for the dwarf galaxies M33 \citep[SIGNALS:][]{Rousseau19} and for the Magellanic Clouds \citep[Local Volume Mapper:][]{Drory24}, which sample substantially different properties than those found in a massive system like \ngc253.
Additionally, MUSE observed a few nearby galaxies at high resolution, including \ngc7793 \citep{Dellabruna20}, M83 \citep{Dellabruna22}, and \ngc300 \citep{McLeod21}, but with significantly smaller coverage compared to our mosaic of \ngc253.
Therefore, while these datasets have comparable, if not better, spatial resolution, they cannot provide the global view needed to connect small-scale processes with the large-scale properties of galaxies such as gas flows, diffuse ionised gas (DIG), or interactions with pristine gas.

Recently, several facilities (e.g., ALMA, \citealt{Leroy21}, Oakes et al. in prep.; JWST, McClain et al. in prep., GO2987, PI: Leroy, Congiu, Faesi; HST, \citealt[][and GO 17809, PI. D. Thilker]{Dalcanton09}; MeerKat, Karapati et al. in prep., MKT-20159, PI: Sardone) have invested significant amounts of observing time to map the entire star-forming disk of \ngc253 at the highest possible spatial resolution.
In this paper, we present our new MUSE mosaic of \ngc253 offering an unprecedented view of its ionised interstellar medium (ISM).
This compilation of multiwavelength data will allow us to explore a wealth of science objectives, starting from the properties, structure, and composition of the multiphase interstellar medium, to the detailed properties of the stellar populations, and the ejection and reaccretion cycle powering the star-formation in the galaxy, just to name a few.
The mosaic, composed of 103 unique pointings, is the largest contiguous extragalactic mosaic observed by MUSE so far\footnote{The absolute largest MUSE mosaic is the Large Magellanic Cloud one published in \citet{Boyce17} and it covers $1\,\si{deg^2}$ with a filling factor of 25\%. Other large mosaics of Milky Way structures are available in the MUSE archive, although they have not been published yet.}.
It covers the vast majority of the star-forming disk, with an average resolution of \SI{15}{pc} ($\sim$ \SI{0.85}{arcsec} PSF FWHM on average).

Here, we focus on testing and validating the data, performing an analysis of the planetary nebulae luminosity function (PNLF) of the galaxy.
Planetary nebulae (PNe) are the final evolutionary stage of intermediate-mass stars \citep[e.g.,][]{Kippenhahn13}. 
During the post-asymptotic giant branch (AGB) phase, these stars eject their outermost layers, creating a gaseous nebula and exposing their core \citep{Osterbrock06}.
The high temperature of the core ($\sim$\SI{100000}{K}) results in a hard spectrum that ionises the surrounding gas, creating a PN.
For this reason, the (optical) emission spectra of PNe are rich with high ionisation collisionally excited lines and recombination lines \citep{Osterbrock06}.
Among them, the strongest one is the \oiii\, line, which can account for up to 13\% of the total energy emitted by a PN \citep[e.g.,][]{Dopita92,Ciardullo10,Schoenberner10}.
Moreover, the ionising source is small, and it produces a limited number of ionising photons.
As a consequence, the typical size of a PN is quite small, with the largest reaching a diameter of $\sim$ \SI{1}{pc} \citep{Acker92}.

Because of these properties, when observed at extragalactic distances PNe appear as unresolved, \oiii-bright sources, making them relatively easy to identify in early-type galaxies where virtually all \oiii\ emission is produced by PNe.
Moreover, their maximum luminosity has been empirically shown to be constant across galaxies and independent of environment and conditions \citep[e.g.][]{Yao23}, to first approximation.
This property, combined with the ease of identification, makes PNe ideal candidates for standard candles.
In fact, since the late 1980s \citep{Jacoby89, Ciardullo89}, their luminosity function (the PNLF) has been a popular secondary indicator to measure distances of galaxies within \SI{30}{Mpc}.

Despite this popularity, the distance of \ngc253 has been estimated by PNLF only twice, by \citet{Rekola05} and \citet{Jacoby24}.
The distances do not agree with each other ($3.34^{+0.26}_{-0.38}$~Mpc for \citealt{Rekola05} and $5.4^{+0.3}_{-0.6}$~Mpc for \citealt{Jacoby24}), but both are based on limited samples.
In this paper, we use our mosaic to identify a new and larger ($\sim$ 20 times) sample of PNe, build the PNLF for \ngc253, and recover an updated independent estimate of the distance of \ngc253.

In Sec.~\ref{sec:data} of this paper we describe the observations, the data reduction procedure, and the spectral fitting.
Section~\ref{sec:analysis} focuses on the detection, characterisation, and cleaning of the planetary nebulae (PNe) sample, while Sec.~\ref{sec:pnlf} describes the fitting of the PNLF and the comparison with the literature findings.
Finally, in Sec.~\ref{sec:discussion} we discuss our results and in Sec.~\ref{sec:summary} we summarise our results.
Table~\ref{tab:deproj} lists the properties we assumed for this galaxy throughout the paper.

\begin{table}[h!]
    \centering
    \caption{\ngc253 properties assumed throughout the paper.}
    \label{tab:deproj}
    \begin{tabular}{lr}
    \hline
    RA              & 00h 47m 33.1s\tablefootmark{a}\\
    Dec             & $-$25d 17m 18.6s\tablefootmark{a}\\
    Position Angle  & $\ang{52.5}$\tablefootmark{a}       \\ 
    Inclination     & $\ang{76}$\tablefootmark{b}    \\ 
    R$_{25}$        & \SI{13.5}{arcmin} (\SI{13.5}{kpc}\tablefootmark{c}) \\
    \hline
    \end{tabular}
    \tablefoot{
    \tablefoottext{a}{From HyperLEDA \citep{Makarov14}}
    \tablefoottext{b}{From \citet{McCormick13}}
    \tablefoottext{c}{Assuming a distance of \SI{3.5}{Mpc} \citep{Okamoto24}}
    }
\end{table}

\section{Observations and data processing}
\label{sec:data}

\begin{figure*}[!h]
        \centering
    \includegraphics[width=\textwidth]{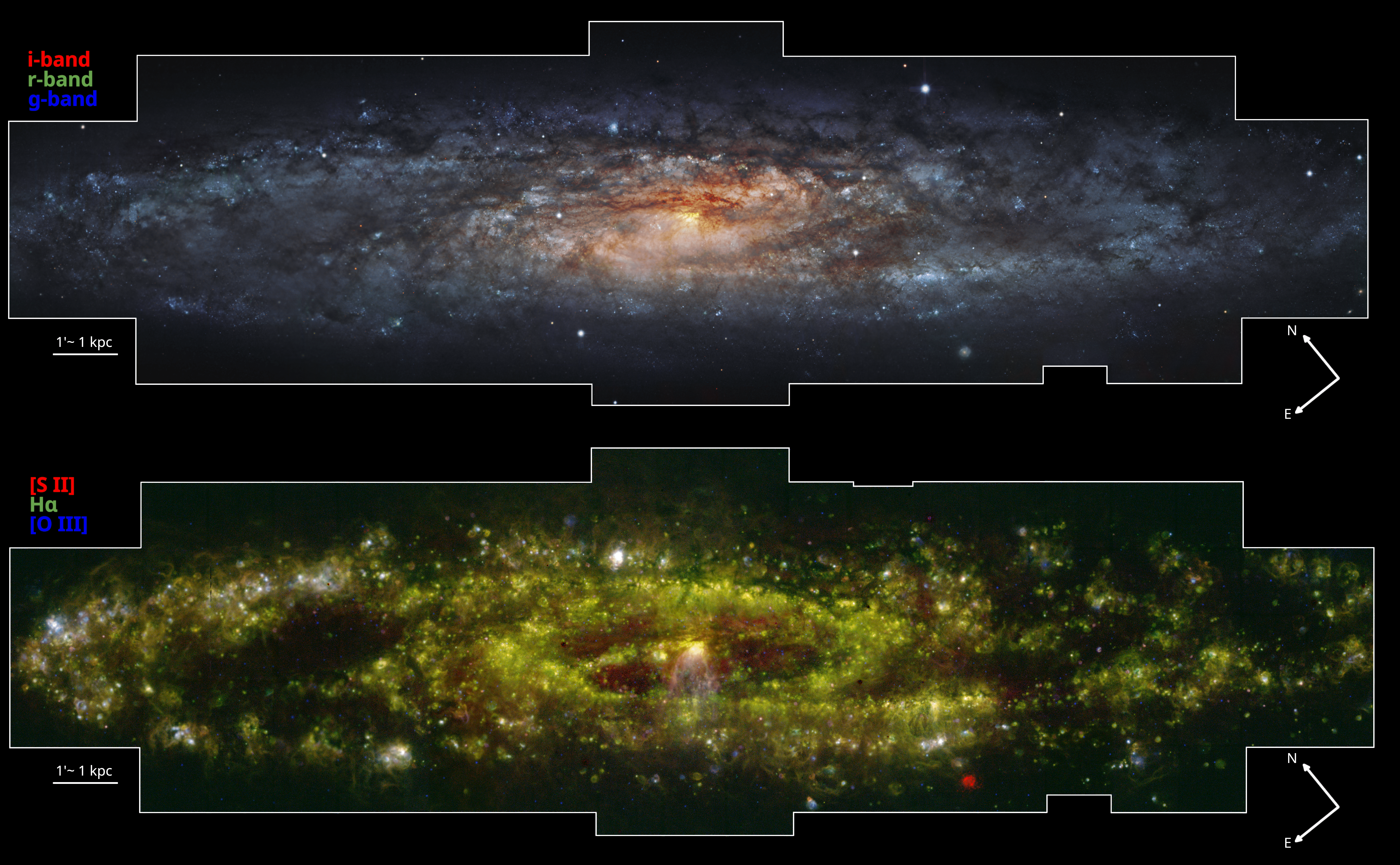}
    \caption{Colour images of \ngc253 produced by combining broad-band images and emission line maps extracted from the MUSE data cube. The mosaic covers an area of $20\times5~\si{arcmin^2}$ and it includes roughly 9 million independent spectra. The top panel show a composition of three broad-band filters: g-band in blue, r-band in green, and i-band in red (Acknowledgement: ESO/M. Kornmesser). We can see the full structure of the galaxy disk, with the prominent bar and complex dust filaments that follow the distribution of the spiral arms of the galaxy. We can also see the change in colour caused by the presence of the central starburst of the galaxy. 
    The bottom panel is a composition of emission line maps with \oiii\ in blue, \Ha\ in green and \sii\ in red. 
    This highlights the multitude of ionised gas structures we observe in the galaxy. 
    The \hii\ regions distribution highlights the structure of the spiral arms, while the \oiii\ and \sii\ emission clearly show the outline of the outflow. Nebulae with different properties can be identified across the field, like the blue, \oiii\ emitting PNe, the green \Ha\ bright \hii\ regions, and the pink, \sii\ emitting SNRs.}
    \label{fig:ngc253_vri}
\end{figure*}

Most of the data presented in this work were acquired as part of ESO programme 108.2289 (P.I. Congiu).
We used MUSE in its seeing-limited wide-field mode configuration and with the extended wavelength range (WFM-NOAO-E), which covers between 4600 and \SI{9300}{\angstrom} with a resolution R$\sim2000$ at \Ha. 
The observations were organised in blocks of 2 pointings with three interleaved \SI{120}{s} sky offset exposures: one at the beginning of each observing block, one between the two pointings, and one at the end.
We observed four \SI{211}{s} exposures per pointing (\SI{844}{s} in total), rotating the field by 90 degrees between each exposure to minimise instrumental signatures in the final mosaic.
We originally planned for 98 pointings, covering an approximate area of $20\times5~\si{arcmin^2}$.
However, three pointings were re-observed because of issues with their positioning, resulting in a final total of 101 pointings and $\sim$51.5 hours of observing time.

The mosaic also includes two archival pointings observed as part of the ESO programme 0102.B-0078 (P.I. Zschaechner).
These data were acquired with MUSE in its seeing-enhancing ground-layer adaptive optics wide-field mode configuration and with the extended wavelength range (WFM-AO-E), covering the same wavelength range as our original data.
Each pointing has been observed with four \SI{490}{s} exposures (\SI{1960}{s} in total), with 90-degree field rotations and small offsets between exposures.
Two \SI{125}{s} sky exposures were also acquired using a typical OSOOSO offset pattern.
The main differences between these archival data and ours are the longer exposure times and a masked spectral region between 5760 and \SI{6010}{\angstrom}, introduced to avoid laser contamination.
Table~\ref{tab:summary_obs} summarises the main properties of the observations, such as date, average airmass, average seeing, and full width at half maximum (FWHM) of the PSF at \SI{5000}{\angstrom}, recovered as described in Sec.~\ref{sec:psf}.

The final mosaic covers an area of $\sim20\times5~\si{arcmin^2}$ ($\sim20\times5~\si{kpc^2}$ at the distance of the galaxy, projected), extending roughly out to $0.75\times R_{25}$.
This corresponds to $\sim$9 million spectra and a final file size of $\sim 300~\si{GB}$.
Figure~\ref{fig:ngc253_vri} shows the reduced mosaic.
The top panel shows a colour image produced by combining g-, r-, and i-band images extracted from the datacube.
The bottom panel highlights the gas emission, showing \oiii, \Ha, and \sii.

\subsection{Data reduction}

All data have been reduced using \verb!pymusepipe!, a python wrapper of the ESO MUSE pipeline \citep{Weilbacher20} presented in \citet{Emsellem22}.
The reduction procedure follows the one described for the PHANGS-MUSE data in the latter work.
However, there are some significant differences that we highlight in the following paragraphs.

\subsubsection{Alignment}
\label{sec:alignment}

The astrometry provided by the MUSE pipeline is not accurate enough for combining multiple exposures into a large mosaic.
\citet{Emsellem22} addressed this by manually aligning each exposure to a reference R-band image from the PHANGS-H$\alpha$ survey (Razza et al. in prep.).
They extracted broad-band images from the MUSE cubes using the PHANGS-H$\alpha$ filter transmission curve, and aligned them by applying sub-pixel shifts and rotations to match the reference image.

While effective, this method is time consuming and somewhat subjective.
We instead used a semi-automatic routine based on the \verb!spacepylot! package\footnote{\url{https://github.com/ejwatkins-astro/spacepylot}}, which employs optical flow to determine alignment corrections.
This technique compares two images of the same object and computes the vector field required to transform one into the other.
\verb!spacepylot! then averages these vectors to derive global shifts and rotations, which we applied to individual exposures following \citet{Emsellem22}.

\ngc253 was not part of the PHANGS-H$\alpha$ survey, but several WFI R-band images are publicly available in the ESO archive.
We reduced them using the pipeline of Razza et al. (in prep.) and \citet{Emsellem22}, and used the resulting R-band mosaic as our astrometric reference.
This image was aligned to Gaia DR2 \citep{GaiaDR2, Lindegren18}.
Comparing stellar positions in the WFI image with Gaia, we found astrometric offsets of $0.02\pm 0.60~\si{arcsec}$ in RA and $0.03\pm0.46~\si{arcsec}$ in Dec.

\subsubsection{Sky subtraction}
\label{sec:skysub}

\begin{figure}
        \centering
    \includegraphics[width=0.49\textwidth]{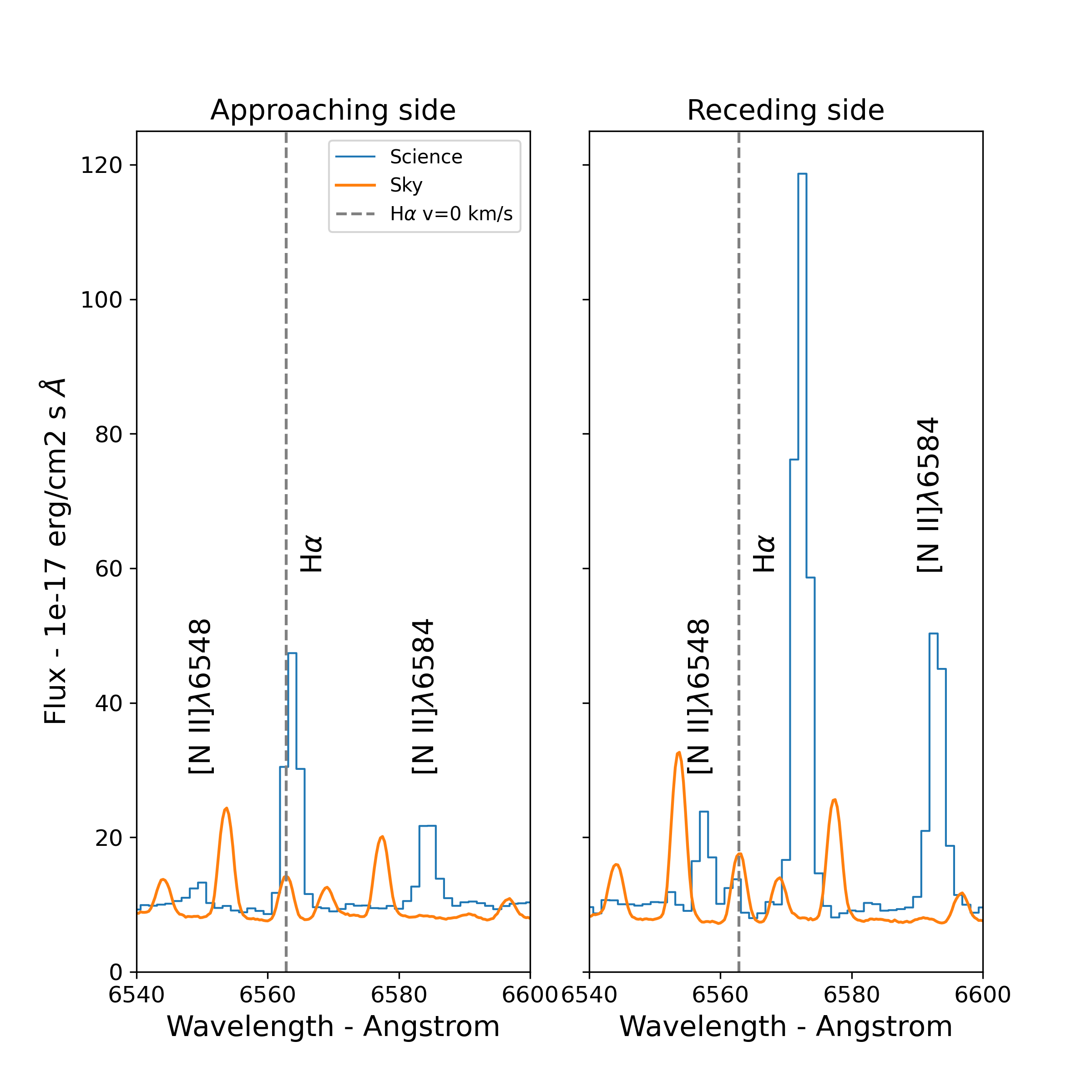}
    \caption{\Ha\ region of the spectra in two different regions of the galaxy. The left panel represent the approaching side of the galaxy, while the right panel the receding side. In both panels we show in blue the spectrum extracted from the science exposure after our updated sky subtraction procedure, in orange the sky spectrum associated with the exposure, while the grey dashed lines represent the expected position of \Ha\ at $v=0~\si{km/s}$.}
    \label{fig:sky_sub}
\end{figure}

\begin{figure}
        \centering
    \includegraphics[width=0.49\textwidth]{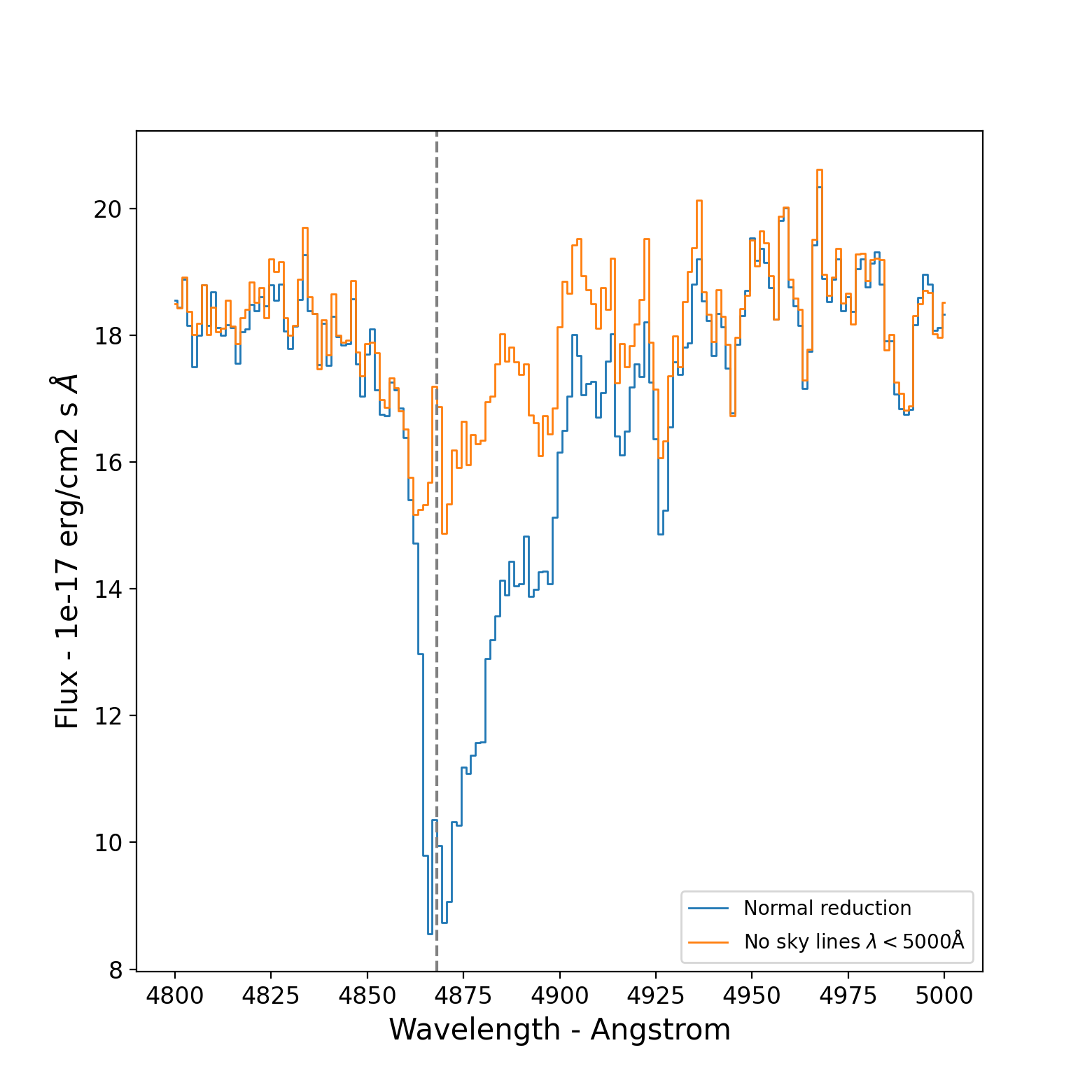}
    \caption{\Hb\ region of the spectra in a \SI{4}{arcsec} circular aperture extracted from one of the pointings with an asymmetric \Hb\ profile. The blue line represents the spectrum obtained from the normal sky subtraction, while the orange line shows the spectrum obtained after removing the sky lines with wavelengths shorter than \SI{5000}{\angstrom} from the line list used to build the sky model from the dedicated sky offset. The vertical, dashed, grey line, represents the expected position of the \Hb\ emission line, considering the rotation curve of the galaxy.}
    \label{fig:hb_prof}
\end{figure}

The sky subtraction is the most critical step in this data reduction.
When observing extended targets with MUSE, the typical strategy is to regularly observe an empty sky field close to the target.
The MUSE pipeline's standard sky subtraction routine\footnote{\verb!skymethod=model! in the \verb!scipost! options.} \citep{Weilbacher20} uses these sky exposures to build a model of the sky emission, separating it into continuum and emission-line components.
The continuum component is directly subtracted from the science exposures.
Emission lines, instead, are grouped by physical origin, compared to the faintest spaxels in the science frames, re-scaled, and then subtracted.
This allows for improved sky subtraction, especially considering that some of these line groups (e.g. [O~\textsc{i}], OH) vary significantly on short timescales and can result in significant artefacts if not treated carefully.

This strategy works well for most extragalactic targets, which typically have a significant redshift.
\ngc253, however, is a very local source, with a systemic line-of-sight velocity of v$_{\rm sys}\sim 242~\si{km.s^{-1}}$ and a rotation curve of $\pm 200~\si{km.s^{-1}}$ \citep{Hlavacek11}.
This means that the approaching side of the galaxy has a relative velocity of only $\sim 40~\si{km.s^{-1}}$, and that some of the target's lines fall very close to their atmospheric counterparts (Fig.~\ref{fig:sky_sub}).
Since this velocity difference is just below the instrument resolution of $\sim 50~\si{km.s^{-1}}$ at \Ha, the pipeline cannot reliably separate sky and target emission, resulting in the over-subtraction of such lines unless they are are not tied to a larger group.
This issue affects mostly the [O~\textsc{i}] lines (especially \oi, [O~\textsc{i}]$\lambda6363$, and [O~\textsc{i}]5577) and the hydrogen Balmer lines (\Ha\ and \Hb).
While the [O~\textsc{i}] lines are among the most variable in the night sky \citep[produced by atmospheric atoms excited through collisions with charged particles;][]{Bates46, Solomon88}, the Balmer lines are more stable \citep[caused by high-atmosphere hydrogen excited by solar Ly$\alpha$;][]{Nossal19}.
Because their intensity varies slowly on the 5--10 minute scale between sky and object exposures, the Balmer lines can be treated as constant during this interval.
This is especially important as they are two of the most important lines in the ionised gas spectrum, and their slow rate of change allow us to recover their flux from the dedicated sky exposures and subtract it directly from the science frames.
To enable this, we excluded \Ha\ and \Hb\ from the pipeline’s list of fitted sky emission lines, forcing them to be treated as part of the continuum.

In addition, we identified several pointings with an asymmetric and unphysical \Hb\ absorption profile (Fig.~\ref{fig:hb_prof}).
This was an artefact caused by the overfitting of sky lines near the \Hb\ region.
Since the Paranal sky spectrum does not contain bright lines at $\lambda < \SI{5000}{\angstrom}$ \citep{Hanuschik03}, we solved the issue by excluding all sky lines bluer than \SI{5000}{\angstrom} from the pipeline's list.
This removed $\sim9000$ lines, almost 40\% of the full list.
Figure~\ref{fig:hb_prof} shows a comparison between the original spectrum (blue) and the corrected one (orange).
Although not all exposures are affected by these issues, we applied the improved sky subtraction procedure to the entire dataset for consistency.
We then combined all reduced exposures into the final mosaic following the method of \citet{Emsellem22}.

\subsection{Spectral fitting}
\label{sec:mosaic}

We processed the data with the PHANGS data analysis pipeline (DAP) to compute the properties and kinematics of the underlying stellar population, subtract their continuum from the spectra, and extract the flux of the main emission lines.
The DAP is described in detail in \citet{Emsellem22}, but here we summarise the main features relevant to this work.
The DAP runs the spectral fitting routine, based on the \verb!ppxf! package \citep{Cappellari17}, three times.
The first run is performed on Voronoi-binned data using a reduced set of simple stellar population models from the E-MILES library \citep{Vazdekis16} to characterise the stellar kinematics across the galaxy.
The second run uses the same Voronoi-binned spectra, keeps the kinematics fixed from the first step, and employs a more extensive library of simple stellar population models \citep[also from the E-MILES library][]{Vazdekis16} to extract additional stellar population parameters (e.g., age, metallicity).
Finally, a third run is performed on each individual spaxel to remove the stellar continuum and simultaneously fit the emission lines with Gaussian profiles.
The kinematics of the stellar templates is fixed from the first step, and the reduced template sample is used for the fit.
Emission lines are grouped into three sets with fixed kinematics \citep[see Sec.~5.2.5 of][]{Emsellem22}.
At the end of the process, the final results are compiled in a multi-extension FITS file, the MAPS file, which contains two-dimensional maps of all fitted parameters.
The physical quantities in the output files are summarised in Tab.~\ref{tab:maps}, while Tab.~\ref{tab:ems} lists the emission lines included in the fit.

We also extract moment-zero, -one, and -two maps of the main emission lines to recover more complete flux and kinematics information in regions where the DAP’s single-Gaussian fit is insufficient to represent the real line profile (e.g., the galaxy centre).
All maps were extracted from continuum-subtracted cubes produced by the DAP using an extraction window of $\pm$\SI{500}{km.s^{-1}} around the recession velocity of the galaxy.
The only exception is the \sii\ doublet, for which we used a larger extraction window ($\pm$\SI{750}{km.s^{-1}}) centred on the average wavelength of the two lines, to avoid cross-contamination.
For these lines, we recovered only the moment-zero map and its associated error map.

\section{Analysis}
\label{sec:analysis}

In this section, we present the analysis we performed to identify, characterise, and clean from contaminants the sample of PNe that we used to build the PNLF and estimate the distance of \ngc253.

\subsection{Detection of the planetary nebulae candidates}
\label{sec:detection}

\begin{figure*}
    \centering
    \includegraphics[width=0.9\textwidth]{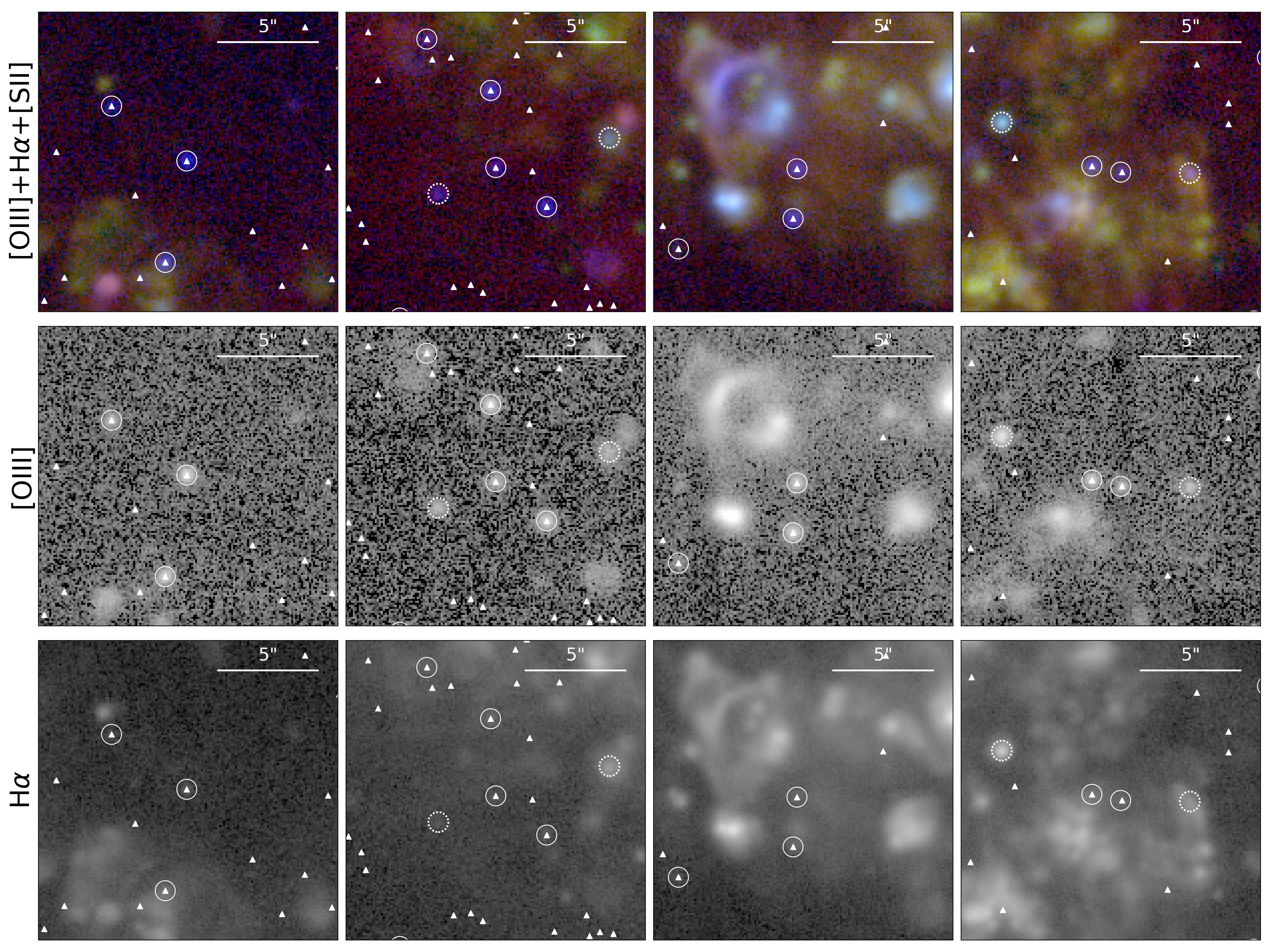}
    \caption{Example of the different images used for the visual detection of nebulae. We show zoom-ins of four different areas of the galaxy. For each region, we show on the top the three colour map (same colours as in Fig~\ref{fig:ngc253_vri}, bottom panel), in the middle the \oiii\, and in the bottom the \Ha\ map. The circles highlight the position of nebulae in the final catalogue, with the solid circles showing the position of confirmed PNe and dotted ones showing nebulae classified either as supernova remnants or as compact \hii\ regions. White triangles show the position of the objects identified by {\tt DAOFIND}, which are, for the most part, spurious detections. }
    \label{fig:det_plot}
\end{figure*}

\begin{table}[]
    \centering
    \caption{Summary of the PNe candidates identified in the paper.}
    \label{tab:sample}
    \begin{tabular}{crrrr}
    \hline\hline
    Sample& Triple & Double & Single & Total\\
    \hline
    E.C.  & 444 & 105 & 42  & 591  \\ 
    C.T.  & 444 & 95  & 163 & 702  \\ 
    T.K.  & 444 & 74  & 46  & 564  \\
    \hline
    Final & 444 & 137 & 251 & 832  \\
    \hline
    \end{tabular}
    \tablefoot{The columns show: the name of the sample (indicated by the initials of the person who created the sample for the original catalogues), the number of objects in common to the three catalogues, the number of objects included in two catalogues, the number of objects identified by a single catalogue, and the number of objects in total.
       }
\end{table}

PNe in galaxies outside the local group appear as \oiii-bright point sources due to their small physical size \citep[$<$\SI{1}{pc}, e.g.,][]{Acker92} and the high degree of ionisation of their gas.  
As a result, all common PNe identification techniques, such as image ``blinking'' \citep[e.g.,][]{Jacoby89}, automatic point source detection \citep[e.g.,][]{Scheuermann22}, colour-magnitude diagram analysis \citep{Arnaboldi02}, or differential emission-line filtering \citep{Roth21}, focus on analysing \oiii\ emission line maps in various forms.

We detect our PNe candidates using a visual approach based on the \oiii, \Ha, and \sii\ moment maps (Fig.~\ref{fig:det_plot}).  
We used both an RGB image combining the three maps (blue for \oiii, green for \Ha, and red for \sii) and the individual \Ha\ and \oiii\ maps. 
We first selected as PNe candidates all sources in the \oiii\ map that appeared point-like and lacked a bright counterpart in the \Ha\ map.  
The RGB map was then used to clarify borderline cases, such as PNe near bright \hii\ regions or objects with detectable \Ha\ emission but still appearing bluer than typical \hii\ regions.  
The colour map clearly highlights PNe as blue dots even in crowded areas or when their emission is faint and hard to identify in the \oiii\ map.  
Ambiguous sources were included in the catalogue, relying on the methods in Sec.~\ref{sec:cleaning} for potential rejection via a quantitative approach.  
We removed from the catalogue all sources for which reliable \oiii\ photometry was not possible, either due to proximity to other \oiii-emitting sources (mainly other PNe candidates) or location at the edge of the mosaic.

We also tested an alternative detection method based on \verb!DAOFIND! \citep{Stetson87}, successfully used by other authors under similar conditions \citep[e.g.,][]{Scheuermann22}.  
This algorithm is quick and semi-automatic but assumes a uniform PSF across the FOV, which is not the case here (see Sec.~\ref{sec:psf}), and cannot exploit colour information as effectively as visual inspection.  
We applied the \verb!photutils! \citep{Photutils24} implementation of \verb!DAOFIND! to the \oiii\ moment-zero map to generate an alternative PNe candidates list.  
We tried several combinations of the main parameters (thresholds, FWHM, sharpness, and roundness), but none yielded satisfactory results.  
The software was detecting $\sim 10000$ objects, which visual inspection showed to be spurious detections or bright spots clearly associated with \hii\ regions.  
Even after removing contaminants (as described in Sec.~\ref{sec:cleaning}), more than half of these sources were still identified as PNe.  
We therefore opted to continue the analysis using the visually compiled catalogue.  
Appendix~\ref{app:daofind} discusses the results of the PNLF using the \verb!DAOFIND!-based catalogue.

However, visual detection can be subjective, and the final candidate list may vary depending on who performs the selection.  
To reduce this bias, three co-authors (E.C., C.T., and T.K.) independently carried out the detection procedure.  
Their catalogues were merged into a single compilation.  
Duplicates were removed by comparing coordinates, merging sources close enough to likely be the same object.  
We then conducted a second check for sources near each other that might cross-contaminate their photometry.  
The final catalogue includes 832 PNe candidates.  
Table~\ref{tab:sample} summarises its composition and that of each person's original sample.

\subsection{Point spread function}
\label{sec:psf}

To obtain reliable photometry of our sources, we need a good characterisation of the point spread function (PSF) of the data.  
It is essential to both select an adequate aperture for the photometry and apply a precise aperture correction.  
However, \ngc253 observations were performed using $\sim53$ OBs (including archival ones) that were acquired between 2018 and 2022 under a wide variety of sky conditions (Table~\ref{tab:summary_obs}).  
As a result, the PSF can change significantly across the mosaic.

A common way to characterise the PSF of astronomical data is to fit bright sources across the FOV of the instrument with a model of the PSF.  
Unfortunately, only a handful of OBs included bright foreground stars that could be used to reliably characterise it, so we devised an alternative method.
Given their properties, PNe are ideal for characterising the local PSF.  
They are faint, but the shape of the MUSE-WFM PSF is relatively well studied, and this ``a priori'' knowledge can be used to decrease the number of free parameters and improve fit reliability.  
In particular, the PSF can be modelled quite well by a circular Moffat profile \citep{Moffat69}, characterised, to the first approximation, by a constant power index \citep[2.8 for the NOAO mode and 2.3 for the AO mode][]{Hartke20, Emsellem22} and variable full width at half maximum (FWHM).

We fitted each PN candidate identified in Sec.~\ref{sec:analysis} with a circular Moffat profile with fixed power index and recovered their FWHM, which we include in the catalogue of PNe candidates available through Vizier.  
The fit was performed using the \oiii\ moment zero map.  
In the following analysis, we consider their FWHM as the FWHM of the local PSF at $\sim$\SI{5007}{\angstrom}, the wavelength of the \oiii\ emission line.

Finally, we take the clean sample of PNe described in Sec.~\ref{sec:cleaning} to estimate the average FWHM of each pointing in which at least one confirmed PNe is present.  
We used the clean sample to ensure only confirmed point sources are considered.  
We report the estimates in Tab.~\ref{tab:summary_obs}.  
The average PSF FWHM of our observations is therefore \SI{0.8}{arcsec} (measured from 100 individual pointings), which corresponds to \SI{13.5}{pc}, with a standard deviation of \SI{0.2}{arcsec}.  
The maximum FWHM is \SI{1.6}{arcsec} (\SI{27}{pc}) and the minimum is \SI{0.49}{arcsec} (\SI{8.5}{pc)}.

\subsection{Aperture photometry}
\label{sec:phot}

We use \verb!photutils! to perform aperture photometry of our candidates using moment-zero maps of several emission lines: \oiii\ to build the PNLF, and \Ha, \nii, and \sii\ to reject contaminants (\hii\ regions, SNRs) in the sample.  
We adopted apertures with diameters equal to the FWHM of the local PSF (see Sec.~\ref{sec:psf}).  
Then, we estimate the local background emission in an annulus with an inner diameter of $4\times$FWHM and an area of five times that of the central aperture.  
To remove the flux component caused by the diffuse emission of the galaxy, we rescaled this background for the difference in size between apertures and subtracted it from the integrated fluxes of the objects.  
Since the apertures we adopted are small, we leveraged the FWHM of the local PSF to calculate the appropriate aperture correction for each PN\footnote{We assume the aperture correction reported in Eq.~3 of \citet{Scheuermann22}.}.  
The PSF fit was performed on the \oiii\ moment-zero map, so we adjust the correction according to the wavelength of the considered line to account for the PSF size's wavelength dependence \citep[as described in][]{Emsellem22}.

We then corrected for Galactic extinction, using a \citet{Cardelli89} extinction law with $R(V) = 3.1$ and an \ebv\ of $0.0165$ \citep{Schlafly11}.  
We did not correct for internal extinction, for the reasons we discuss in Sec.~\ref{sec:extinction}.  
Finally, we converted the \oiii\ fluxes (in $\si{erg.cm^{-2}.s^{-1}}$) to apparent magnitudes using the following equation from \citet{Jacoby89}:  
\begin{equation}
\label{eq:o3mag}
    m_{5007} = -2.5 \cdot \log_{10} I_{5007} - 13.74\;.  
\end{equation}

We estimated errors on the integrated fluxes using the moment-zero error maps and standard error propagation.  
We also add in quadrature a \SI{0.11}{mag} error on the \oiii\ magnitudes, which we estimate as the typical uncertainty on the aperture correction from our analysis in Sec.~\ref{sec:ap_cor}.

\subsection{Contaminants}
\label{sec:cleaning}

The next step is to clean our sample of PNe candidates from contaminants.  
This is particularly important in a star-forming galaxy like \ngc253 where PNe are a minority of the \oiii-emitting sources.  
The two most common contaminants are supernova remnants (SNR) and compact \hii\ regions.  
To remove them from our catalogue, we leverage the different spectral properties of the three classes of sources.  
\hii\ regions and PNe are both powered by photoionisation, but the former are ionised by O or B stars that produce a soft ionising spectrum with respect to the white dwarfs typically powering PNe.  
For \hii\ regions, this results in a spectrum dominated by hydrogen recombination lines (\Ha\ in particular) and other lines from moderately low ionisation ions (e.g., \niion).  
Planetary nebulae spectra, on the other hand, are dominated by highly ionised forbidden lines, like \oiii, and recombination lines, such as \heii. 
Finally, supernova remnants are mostly powered by shocks, which enhance low-ionisation lines like the \oi\ and \sii\ doublets.

Several methods have been developed in the literature to take advantage of these properties to classify these nebulae.  
Here, we exploit the two most common.  
To reject \hii\ regions, we leverage the criterion developed by \citet{Ciardullo02} and \citet{Herrmann08}:  
\begin{equation}
\label{eq:HIIregion_criteria}
\log_{10} \frac{I_{5007}}{I_{\mathrm{H}\alpha+[\mathrm{N}\,\textsc{ii}]\,\lambda6583}} > -0.37 M_{5007} - 1.16 \;,
\end{equation}
while for SNRs we apply the traditional \citet{Dodorico80} criterion:  
\begin{equation}\label{eq:SNR_criteria}
 \log_{10} \frac{I_{[\mathrm{S}\,\textsc{ii}]\,\lambda6717 + [\mathrm{S}\,\textsc{ii}]\,\lambda6731}}{I_{\mathrm{H}\alpha}} > -0.4\;.
\end{equation}
\citet{Kopsacheili20} and \citet{Li24} showed how an \oi\ based criterion is more effective at identifying SNRs from \hii\ regions.  
However, the \oi\ line in our data is heavily compromised by sky emission and we are not interested in a clean sample of SNRs, so we stuck with the criterion in Eq.~\ref{eq:HIIregion_criteria}.  
During the classification, we consider the flux of non-detected lines equal to its uncertainty.  
We also checked for overluminous sources, that is, sources that look like PNe but that are significantly brighter than the bright cutoff of the PNLF \citep{Longobardi13,Hartke17,Roth21,Scheuermann22}, without finding any of them in our sample.  
Finally, we exploited the size recovered in the previous section, to reject five objects that are either too large (FWHM $>$ \SI{2}{arcsec}) or too small (FWHM $<$ \SI{0.2}{arcsec}).  
The first criterion rejects extended sources, since all our data were acquired with seeing $\sim$ \SI{1.2}{arcsec} or better.  
The second one rejects bad pixels and similar artefacts that can be confused with real sources at first glance.

\begin{figure*}
    \centering
    \includegraphics[width=1\textwidth]{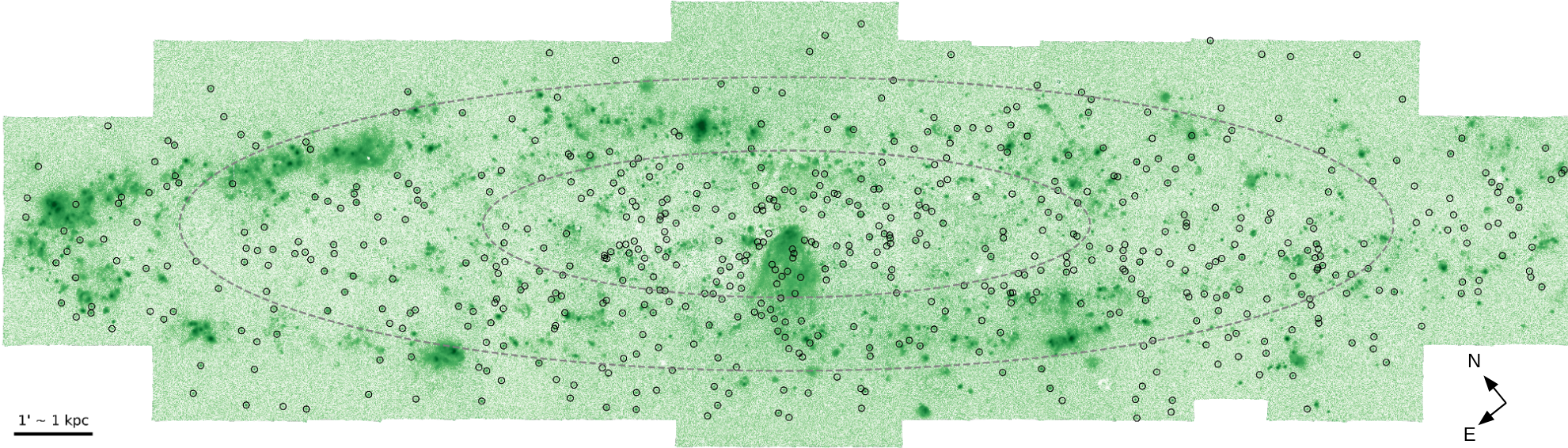}
    \caption{\oiii\ line map of \ngc253. Empty circles mark the location of confirmed PNe. The grey dashed ellipses represent the boundary used to defined the radial bins presented in Sec.~\ref{sec:radial}.}
    \label{fig:OIII_map_PNe}
\end{figure*}

After all these checks, our clean sample of PNe includes 571 objects.  
Figure~\ref{fig:OIII_map_PNe} shows the position of the confirmed PNe in the \oiii\ emission line map.  
Of the remaining 253 regions (without counting the 5 regions rejected for their size), 200 are classified as \hii\ regions, while 53 as SNRs.  
The catalogue of sources, which contains all the information reported in Table ~\ref{tab:cat_example}, is available through CADC and Vizier.

\section{The planetary nebula luminosity function in \ngc253}
\label{sec:pnlf}

The PNLF, that is, the number of PNe observed as a function of their \oiii\ luminosity, is commonly described by an empirical relation, an exponential distribution truncated at the bright end \citep{Ciardullo89}:
\begin{equation}
    N (M_{5007}) \propto \mathrm{e}^{0.307\, M_{5007}} \left(1 - \mathrm{e}^{3\left(M^{*}_{5007}- M_{5007}\right)}\right)~,
	\label{eq:pnlf}
\end{equation}
where $M^*_{5007}$ denotes the absolute \oiii\ magnitude of the brightest possible PN, i.e., the zero-point of the luminosity function. 
Its value is calibrated using galaxies with distances known from primary indicators such as Cepheids or the Tip of the Red Giant Branch (TRGB).  

Typical estimates place $M^*_{5007}$ around \SI{-4.5}{mag}, with $-4.54$~mag from \citet{Ciardullo13} being the most widely adopted.  
While a mild metallicity dependence has been reported \citep[e.g.,][]{Ciardullo02, Bhattacharya21, Scheuermann22}, its significance remains uncertain.  
Given that \ngc253 has near-solar metallicity \citep[$12+\mathrm{log(O/H)} = 8.69$;][]{Beck22}, where the empirical metallicity dependence is nearly flat, we ignore this dependence and consider $M^*_{5007} = -4.54~\si{mag}$ throughout our analysis.  

\begin{figure*}
    \centering
    \includegraphics[width=\linewidth]{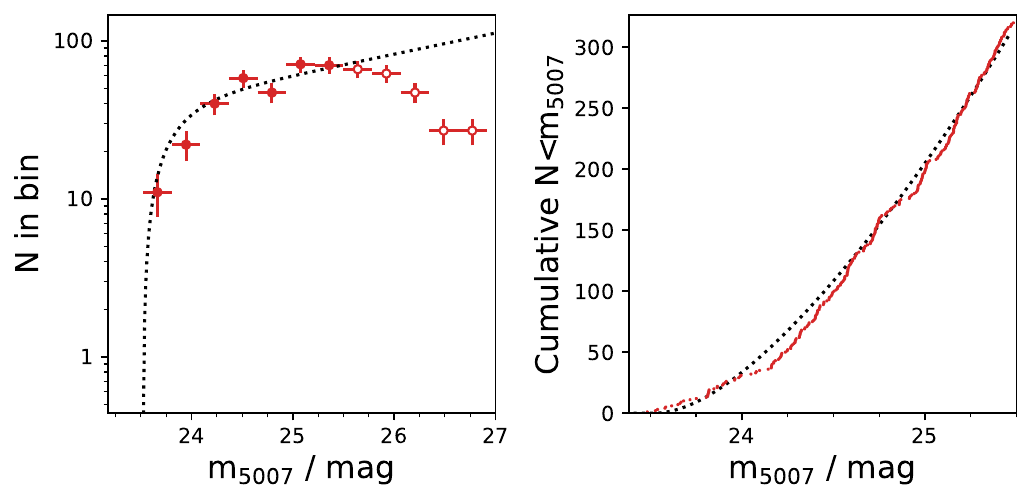}
    \caption{PNLF and cumulative PNe \oiii\ luminosity function for \ngc253. The red points show the measured PNLF and cumulative function, while the black dashed line represents the best fit model. Empty points marks the bins of the PNLF not considered in the fit.}
    \label{fig:pnlf}
\end{figure*}

\begin{figure*}
    \centering
    \includegraphics[width=\linewidth]{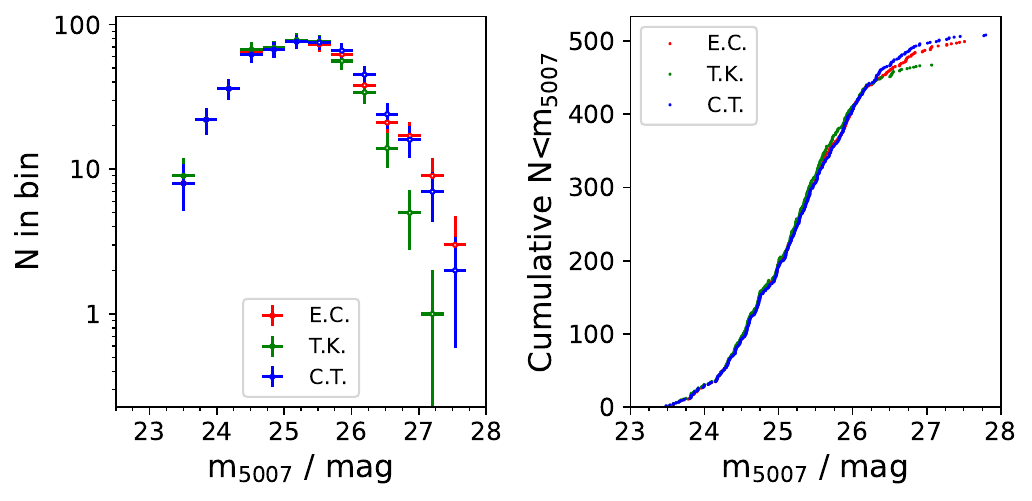}
    \caption{Comparison between the luminosity functions obtained from the three individual samples described in Sec.~\ref{sec:detection}. The left panel shows the PNLFs, while the right panel shows the cumulative luminosity function. The figure shows how the three samples significantly overlap at the bright end, while the number of PNe drops in all cases around \SI{25.5}{mag}.}
    \label{fig:completeness}
\end{figure*}

We fit Eq.~\ref{eq:pnlf} to our data using the maximum likelihood approach of \citet{Scheuermann22}.  
Figure~\ref{fig:pnlf} shows the resulting PNLF (left) and its cumulative form (right).  
To avoid incompleteness effects at the faint end, we impose a magnitude cutoff based on visual comparison of three independently compiled PN samples (E.C., T.K., C.T.).  
Figure~\ref{fig:completeness} shows that all three samples begin to drop at $\sim$\SI{25.5}{mag}.
While a real decline in PN counts is expected $\sim$2~mag below the bright end \citep[e.g.,][]{Jacoby02, Reid10, Ciardullo10, Rodriguez15, Bhattacharya21}, our model does not account for this feature.  
We therefore restrict the fit to the 320 PNe brighter than \SI{25.5}{mag}. 
The fit returns a distance modulus of $28.07^{+0.04}_{-0.05}$~mag ($4.10^{+0.07}_{-0.09}$~Mpc).

\begin{figure}
    \centering
    \includegraphics[width=0.5\textwidth]{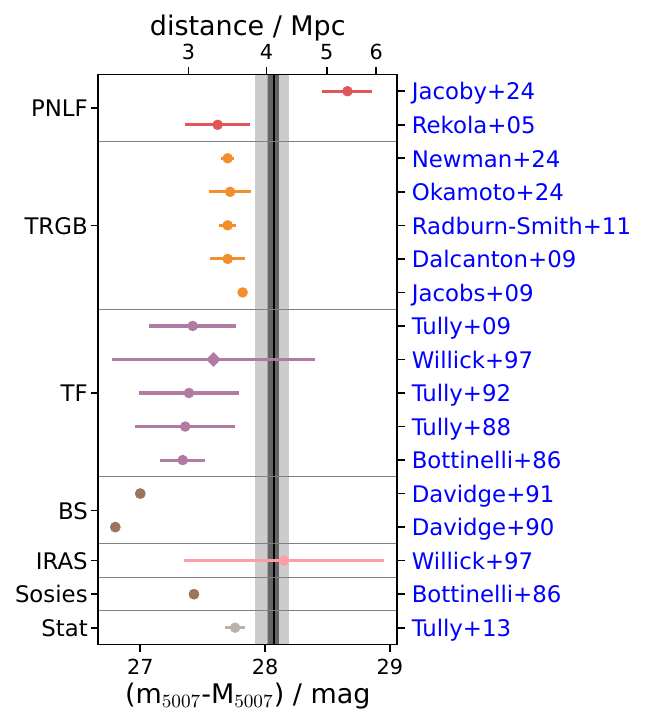}
    \caption{Comparison between our measurement (solid black vertical line) with errors ($1\sigma$, dark colour, and $3\sigma$, lighter colour) with distances from the literature. All measurements are divided depending on the method used. We use the following abbreviations: Planetary Nebula Luminosity Function (PNLF), Tip of the Red Giant Branch (TRGB), Tully-Fisher (TF), Brightest Stars (BS), Infra-Red Astronomical Satellite (IRAS), Sosies, and Statistical (Stat). Distances from: \citet{Bottinelli86,Tully88,Davidge90,Davidge91,Tully92,Willick97,Rekola05,Dalcanton09,Jacobs09,Tully09,RadburnSmith11,Tully13,Jacoby24,Newman24,Okamoto24}.}
    \label{fig:distances}
\end{figure}

Figure~\ref{fig:distances} compares the distance from our analysis to previous results from the literature.  
Our measurement, shown as a solid black vertical line, is accompanied by shaded regions indicating the $1\sigma$ (dark grey) and $3\sigma$ (light grey) uncertainties. 
It stands out as significantly larger than most previously reported values.  
The commonly accepted distance to NGC 253 is $\sim$~\SI{3.5}{Mpc}, with the most recent estimates reported by \citet{Newman24} and \citet{Okamoto24} via TRGB analysis.  
Our measurement exceeds this value by $\sim$\SI{0.35}{mag} (\SI{0.6}{Mpc}), with only two similar distances found in \citet{Willick97}, based on IRAS satellite data and the Tully-Fisher relation \citep{Tully77}.
Among the compiled distances, we have two other measurements via the PNLF: \citet{Rekola05} and \citet{Jacoby24}.  
In the following, we will directly compare our result with those two studies.

\begin{figure}
    \centering
    \includegraphics[width=0.5\textwidth]{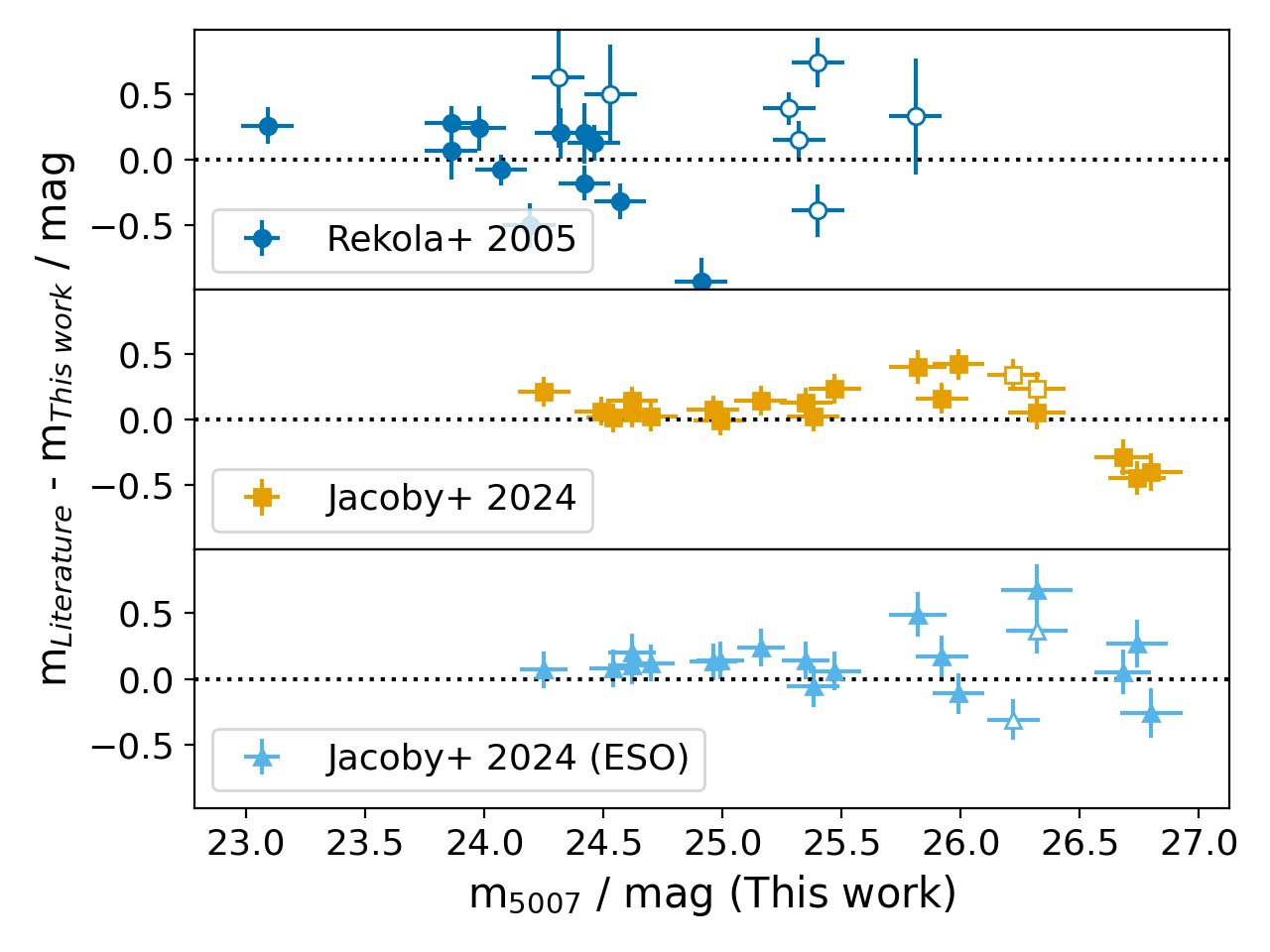}
    \caption{Comparison between our PNe photometry, \cite{Rekola05} and \cite{Jacoby24}. For \citet{Jacoby24} we performed the comparison using both the data reduced by us, and the cubes directly recovered from the ESO archive. Solid symbols represent PNe that were used by the original authors in their PNLF computation. Open symbols represent objects that were not included in the final fit.}
    \label{fig:phot_comp}
\end{figure}

\subsection{Comparison with \citet{Rekola05}}
\label{sec:rekola}

\citet{Rekola05} were the first to measure the distance of \ngc253 using the PNLF method through ground-based \oiii\ and \Ha\ narrow-band images acquired with FORS.  
Of the 24 PNe they detected, they selected the 14 brightest ones for the PNLF fit and obtained a distance of $27.62^{+0.16}_{-0.26}$~mag ($3.34^{+0.26}_{-0.38}$~Mpc).  
Of these 14 PNe, 12 have counterparts in our sample\footnote{Of the two missing objects the first is outside of our mosaic, while the second is not in our sample because of its extended morphology in the MUSE data.}.  
We classify all of them as PNe, except one, the brightest in the \citet{Rekola05} sample, that we classify as an \hii\ region.  

The blue circles in Fig.~\ref{fig:phot_comp} show the difference between our photometry and \citet{Rekola05}, with solid markers denoting the sources used in their PNLF fit and open circles indicating the ones they excluded but that are within our mosaic.  
Their Milky Way extinction correction slightly differs from ours.  
To ensure consistency in the comparison, we recalculated our extinction correction using $\ebv~=0.019$, R$_V=3.07$, and a \citet{Fitzpatrick99} extinction law.  
There is a scatter of \SI{0.4}{mag} between the two works, but without a significant offset (\SI{0.06}{mag}).  
This implies that there is no systematic difference between the two sets of measurements.  
The observed scatter may be related to the different observing techniques and the inherent challenge of obtaining precise photometry for dim objects against a complex background emission, such as that present in \ngc253.  

To confirm that the difference in photometry does not influence the distance of the galaxy, we perform the fit of the PNLF using only the 12 sources in common between the two samples.  
We measure a distance modulus of $27.38^{+0.14}_{-0.30}$~mag ($3.00^{+0.19}_{-0.42}$~Mpc).  
This value is slightly smaller, but still within the error bars of \citet{Rekola05} original measurement, confirming that the difference in photometry is not the cause of the discrepancy. 
However, we highlight that we include in this fit the object we classify as an \hii\ region.  
Given the small number of PNe, the PNLF distance is particularly sensitive to the magnitude of the brightest objects, and this nebula is significantly brighter than the second brightest one in both samples (\SI{0.34}{mag} according to \citealt{Rekola05} photometry and \SI{0.77}{mag} according to our photometry).  
Therefore, we repeated the analysis removing this object from the sample, obtaining a distance modulus of $28.02^{+0.12}_{-0.29}$~mag ($4.01^{+0.21}_{-0.54}$~Mpc).  
Although the errors are large, the value is consistent with what we recover from the full sample, confirming that the discrepancy between our results and those reported in \citet{Rekola05} is mainly driven by the misclassification of this single object.

\subsection{Comparison with \citet{Jacoby24}}
\label{sec:jacoby}

\citet{Jacoby24} recently estimated the distance of \ngc253 via the PNLF using the two archival MUSE pointings from programme 0102.B-0078 (P.I. Zschaechner).  
Their analysis employs the differential emission line filter method introduced by \citet{Roth21} to identify PNe candidates. 
This method comprises two steps: first, generating continuum-subtracted maps for each spectral channel within a specific wavelength range centred on the \oiii\ line; second, detecting PNe candidates by combining three wavelength-adjacent maps and selecting point sources that appear consistently across at least three of them.  
Despite the small area covered by the data, they were able to identify 34 PNe and estimate a distance modulus of $28.66^{+0.12}_{-0.28}$~mag ($5.4^{+0.3}_{-0.6}$~Mpc).  
This is one magnitude ($\sim \SI{2}{Mpc}$) larger than the typically accepted value.  
They note that the archival MUSE pointings cover a suboptimal region for PNLF studies, the galaxy's centre, where gas and dust structures complicate PNe identification and hinder precise photometry, likely leading to an overestimated distance.  
Nevertheless, we proceed with the comparison since it could still provide an important validation of our results.  

We have included the same MUSE archival data in our mosaic, so all PNe identified by \citet{Jacoby24} are included in our footprint.  
We identify in our catalogue only 21 of their 34 objects, and all of them are classified as PNe, except one (which we consider a SNR).  
Visual inspection suggests that most of the remaining 13 sources are either extremely faint in our \oiii\ map or appear as knots within the outflow-related \oiii\ emission.  
One exception is a source at the edge of the southern pointing, excluded from our catalogue because it appears as a pair of close candidates, making reliable photometry unfeasible.  
The orange points in Fig.~\ref{fig:phot_comp} illustrate the photometric comparison between the two works.  
The scatter (\SI{0.17}{mag}) and offset (\SI{0.07}{mag}) are similar to those found in the comparison with \citet{Rekola05}, though the residuals reveal some structured trend, hinting at possible systematic differences.  

\citet{Jacoby24} used for their analysis the reduced version of the data publicly available through the ESO archive, which uses a different approach to the sky subtraction and exposure alignment.  
To assess whether this affects the measurements, we repeated our analysis using these publicly available cubes and found no significant difference for the brightest sources (Fig.~\ref{fig:phot_comp}, light blue triangles).  

Following the same approach used for \citet{Rekola05}, we attempted to reproduce the \citet{Jacoby24} PNLF using only the sources common to both catalogues\footnote{Here we consider our original, best-estimate photometry.}.  
Using all the sources we have in common, we obtain a distance of $28.47^{+0.14}_{-0.31}$~mag ($4.95^{+0.32}_{-0.70}$~Mpc).  
If we remove the two sources that were not used by \citet{Jacoby24} in their fit, the result changes to $28.44^{+0.15}_{-0.32}$~mag ($4.88^{+0.34}_{-0.71}$~Mpc).  

Our photometry yields generally fainter magnitudes than those in \citet{Jacoby24}, leading to a slightly smaller distance estimate.  
However, the errors are quite large, so the result is still within $1\sigma$ of their measurement.  
Moreover, this distance is significantly larger than what we obtain using our full PNe sample, supporting the interpretation by \citet{Jacoby24} that the brightest PNe in the galaxy centre are likely missing.

\paragraph{}
This exercise demonstrates two things.  
First, it shows that, even though our photometry shows some difference with respect to other measurements available in the literature, this does not  significantly influence the distances recovered by the PNLF when limiting our analysis to the sample in common with previous works, and that therefore our result is reliable.  
Secondly, it shows that the \citet{Rekola05} estimate of the distance of the galaxy, the only one in agreement with the commonly accepted TRGB-based value, is driven by the misclassification of one object.
Therefore, all PNLF-based measurements of the distance to this galaxy are significantly overestimated compared to the TRGB-based distances.  
In the next section, we will investigate the possible origin of this tension.

\section{Discussion}
\label{sec:discussion}

\begin{figure}
    \centering
    \includegraphics[width=\linewidth]{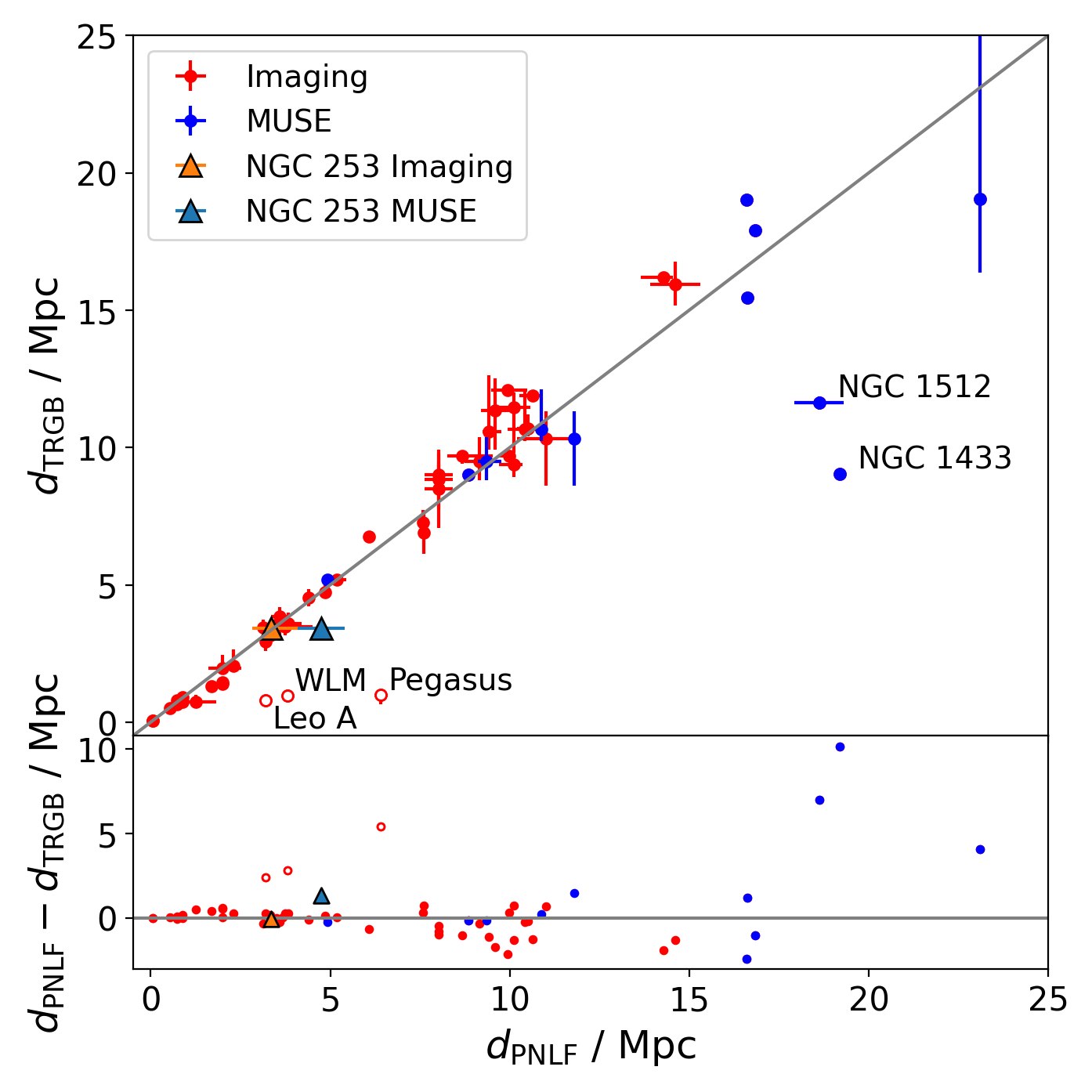}
    \caption[]{Distance comparison for galaxies with both PNLF and TRGB measurements from NED\footnotemark. Red points indicate galaxies where the PNLF was measured using imaging techniques, while blue points indicate PNLF measurements obtained with MUSE data. The grey line shows the one-to-one relation. The orange and light blue triangles identify the position of NGC 253 in the diagram when considering only imaging PNLF or MUSE-based PNLF respectively. Error bars represent the range between the minimum and maximum values obtained via a specific method. We also annotated in the main panel the position of a few notable sources, and we plot Pegasus, Leo A, and WLM with open red circles.}
    \label{fig:pnlf_trgb}
\end{figure}

As we introduced in Sec.~\ref{sec:pnlf}, the distance modulus we obtain from our analysis is \SI{0.35}{mag} (\SI{0.6}{Mpc}) larger than what is recovered by other methods (mostly TRGB, $\sim$\SI{27.72}{mag}, $\sim$\SI{3.5}{Mpc}, \citealt{Jacobs09, Dalcanton09, RadburnSmith11, Okamoto24, Newman24}).  
Given the good reported agreement between these two methods in past work \citep[e.g.][]{Roth21, Scheuermann22, Jacoby24}, this discrepancy was unexpected.  
Figure~\ref{fig:pnlf_trgb} shows a direct comparison between TRGB and PNLF distances recovered from the NASA Extragalactic Database (NED) Redshift-Independent Distances database\footnotetext{\url{https://ned.ipac.caltech.edu/Library/Distances/}}\footnotemark[\value{footnote}].  
Since the database was last updated in 2020, it does not include more recent MUSE-based PNLF measurements.  
We therefore supplement the plot with additional distances published in \citet{Anand21}, \citet{Roth21}, \citet{Scheuermann22}, \citet{Jacoby24}, and \citet{Anand24}.  
The figure illustrates that TRGB- and PNLF-based distances follow a tight relation out to $\sim$\SI{15}{Mpc}.\footnote{ 
The three points that significantly deviate from the relation at low distances represent very nearby dwarf galaxies: Pegasus, Leo A and WLM. 
Their PNLF-based distances are early attempts and are based on a very limited sample of PNe \citep[one for Pegasus, one for Leo A and two for WLM;][]{Jacoby81}.}
Therefore, we are not considering these three points in the analysis.  
Beyond \SI{15}{Mpc}, only a handful of MUSE-based PNLF measurements are available, with increased scatter driven primarily by \ngc1433 and \ngc1512.  
As discussed by \citet{Scheuermann22} and \citet{Jacoby24}, these deviations are linked to the misidentification of the TRGB in their colour-magnitude diagrams (CMD).  

In contrast, this explanation does not hold for \ngc253.  
Its distance is significantly lower, making it easier to identify the position of the TRGB in the CMD. 
In fact, its TRGB-based distance has been independently measured multiple times in the literature \citep{Jacobs09, Dalcanton09, RadburnSmith11, Okamoto24, Newman24}, and all estimates are in mutual agreement.  
On the other hand, the PNLF is not an easy method to apply to star-forming galaxies, where most of the \oiii\ emission is produced by other types of nebulae.  
Additionally, there are known cases where the PNLF yields inconsistent distances depending on the region of the galaxy sampled \citep[][]{Herrmann08, Bhattacharya21}.  
Several observational effects, such as extinction, background subtraction, and aperture correction, can impact the photometry of PNe and hence the distance derived from the PNLF.  
In the following section, we examine these factors in more detail to investigate the origin of the discrepancy.

\subsection{Aperture correction}
\label{sec:ap_cor}

\begin{figure}
\centering
    \includegraphics[width=\linewidth]{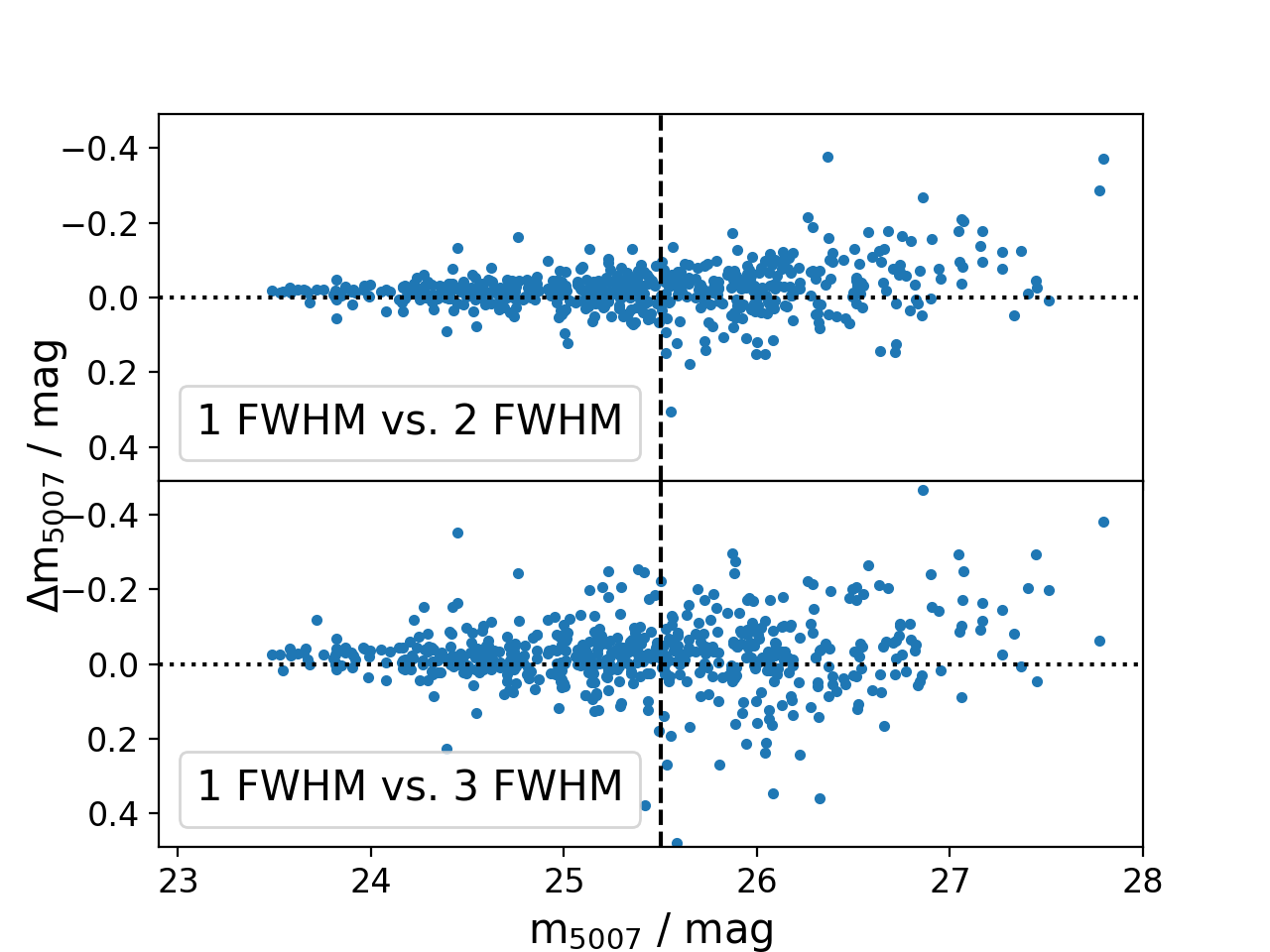}
    \caption{Effect of the aperture correction on the measured magnitude of the PNe in our sample. Only the objects brighter than \SI{25.5}{mag} (vertical dashed line) have been used for the PNLF fit. We define $\Delta\mathrm{[OIII]}$ as $m(X\times \mathrm{FWHM}) - m(\mathrm{FWHM})$ with $X$ being either 2 or 3.}
    \label{fig:ap_corr}
\end{figure}

The aperture correction is one of the most critical aspects of our analysis.
It is essential to recover the full flux of a point-like source when a small aperture, which does not encompass the entirety of the PSF, is used to perform photometry.
Good aperture corrections should provide the same total flux for the same object regardless of the aperture size.

To assess whether the aperture correction could account for the discrepancy in our distance estimate, we conducted a series of validation tests.  
First, we repeated the photometry using apertures with diameters 2 and 3 times larger than the originals, and applying the corresponding aperture corrections.
We then compared the resulting fluxes to our original measurements.  
We evaluated that doubling the diameter results in fluxes that are, on average, only 2\% higher, with a standard deviation of 6\%.  
Instead, tripling the diameter produces fluxes that are 3\% higher on average, but with a slightly larger standard deviation (12\%).  
The small average differences indicate that the aperture correction performs well overall, although the increased scatter suggests some object-to-object variability.  
However, Fig.~\ref{fig:ap_corr} shows that the scatter is negligible for the brightest objects, which means that the effect of the aperture correction on the distance estimate is also negligible.  

To confirm this, we used the new photometry to recompute the galaxy's distance.  
We obtain a distance modulus of $28.05^{+0.04}_{-0.05}$~mag ($4.07^{+0.07}_{-0.09}$~Mpc) when using the $2\times$FWHM apertures and $28.02^{+0.04}_{-0.05}$~mag ($4.01^{+0.07}_{-0.09}$~Mpc) when using the $3\times$FWHM apertures.  
The final result is, in both cases, within $1\sigma$ of our original measurement ($28.07^{+0.03}_{-0.05}$~mag or $4.10^{+0.07}_{-0.09}$~Mpc).  
We therefore conclude that, although our aperture correction method introduces some scatter, its effect on the final distance modulus is minimal and limited to at most \SI{0.05}{mag}. 

\subsection{Radial variation of the PNLF}
\label{sec:radial}

\begin{figure}
    \centering
    \includegraphics[width=\linewidth]{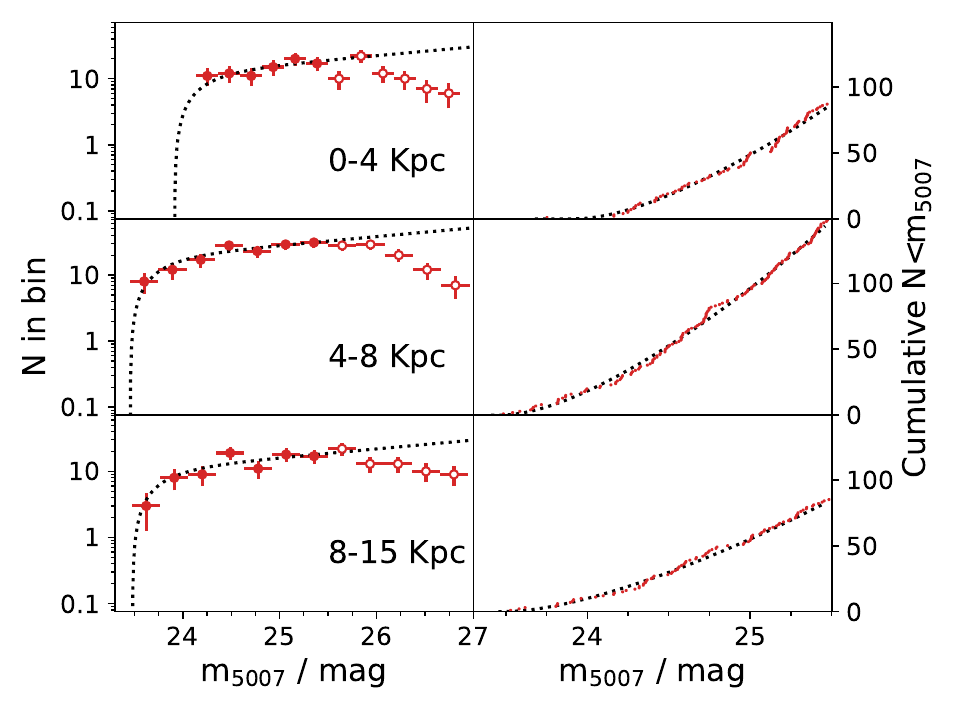}
    \caption{PNLF for the three subsample of PNe at different galactocentric distances. The left column show the PNLF and their fit, while the right column show the cumulative luminosity function. Colours and symbols are the same as described in Fig.~\ref{fig:pnlf}.}
    \label{fig:spatial}
\end{figure}

The PNLF has generally been found to be stable when selecting subsamples of PNe located in different regions of a galaxy \citep[e.g.][]{Hui93, Ciardullo04, Ciardullo13}. 
However, some exceptions have been observed, both in the halo \citep[e.g., M31;][]{Bhattacharya21} and within galaxy disks \citep[e.g.,][]{Herrmann08}.
The comparison we performed in Sect.~\ref{sec:jacoby} suggests that a radial dependence of the PNLF may also be present in \ngc253, as the distance recovered from PNe located in the central regions is significantly larger than that obtained using the full population.

To further explore a potential environmental dependence, we divided our PNe into subsamples based on their deprojected distance from the galaxy centre.
We deproject the position of each object using the parameters listed in Table~\ref{tab:deproj}, which are adopted from the HyperLEDA catalogue \citep{Makarov14} and \citet{McCormick13}, and a standard distance of \SI{3.5}{Mpc} \citep[e.g.,][]{Okamoto24}.
We then defined three subsamples: the first includes PNe located within the central \SI{4}{kpc} of the galaxy; the second contains those between \SI{4}{kpc} and \SI{8}{kpc}; and the third includes all objects beyond \SI{8}{kpc}.
These boundaries were chosen to divide the galaxy’s semi-major axis into roughly equal parts while ensuring that each bin contains a sufficient number of PNe brighter than \SI{25.5}{mag} ($\sim$100) to allow a reliable fit of the PNLF (see Table~\ref{tab:kpc_sample}).

\begingroup

\setlength{\tabcolsep}{10pt} 
\renewcommand{\arraystretch}{1.5} 

\begin{table}[]
    \centering
    \caption{Summary of the three distance based samples and results of the fit. }
    \label{tab:kpc_sample}
    \begin{tabular}{lcccc}
    \hline\hline
    Sample & All & Used& $\mu $ (mag)& D (Mpc)\\
    \hline
    All & 571 & 320 & $28.07^{+0.04}_{-0.05}$ & $4.10^{+0.07}_{-0.09}$\\
    $0$--$\SI{4}{kpc}$ & 158 & 87& $28.46^{+0.06}_{-0.09}$ & $4.91^{+0.14}_{-0.21}$\\
    $4$--$\SI{8}{kpc}$ & 251 & 148& $28.00^{+0.05}_{-0.08}$ & $3.97^{+0.09}_{-0.14}$\\
    $>\SI{8}{kpc}$ & 162 & 85& $28.02^{+0.07}_{-0.10}$ & $4.02^{+0.12}_{-0.19}$\\
    \hline
    \end{tabular}
    \tablefoot{
    The columns show: the name of the sample, the number of PNe contained in the sample, the number of PNe used in the PNLF fit, the distance modulus, the distance in Mpc.
    }
\end{table}

\endgroup

The result of the analysis is reported in Fig.~\ref{fig:spatial}. 
The plot shows that the central PNLF is almost \SI{0.75}{mag} fainter than that obtained from the other two samples. 
As a consequence, the distance is significantly larger ($28.46^{+0.06}_{-0.09}$~mag, or $4.91^{+0.14}_{-0.21}$~Mpc), and it agrees with the value we derived in Sect.~\ref{sec:jacoby}. 
The other two PNLFs appear mutually consistent, although the $4$--$\SI{8}{kpc}$ sample contains a larger number of sources. 
Indeed, the distance moduli we recover are $28.00^{+0.05}_{-0.08}$~mag ($3.97^{+0.09}_{-0.14}$~Mpc) and $28.02^{+0.07}_{-0.10}$~mag ($4.02^{+0.12}_{-0.19}$~Mpc) for the $4$--\SI{8}{kpc} and $>$\SI{8}{kpc} samples respectively.

We performed a k-sample Anderson-Darling test for each pair of subsamples. 
The results are reported in Tab.~\ref{tab:adtest}, and they confirm that the central subsample is statistically inconsistent with being drawn from the same parent distribution as the other two subsamples at high significance levels (1\% and 3\%, respectively), while the two outer subsamples are fully consistent with being drawn from the same distribution.

\begingroup

\setlength{\tabcolsep}{6pt} 
\renewcommand{\arraystretch}{1.5} 

\begin{table}[]
    \centering
    \caption{Results of the K-sample Anderson-Darling test applied to each couple of sub-samples.}
    \label{tab:adtest}
    \begin{tabular}{lccc}
    \hline\hline
    Sample & Statistic & p-value& Significance\\
    \hline
    $0$--$\SI{4}{kpc}$ vs $4$--$\SI{8}{kpc}$& 3.670 & 0.011& 1\%\\
    $0$--$\SI{4}{kpc}$ vs $>\SI{8}{kpc}$ & 2.363 & 0.035& 3\%\\
    $4$--$\SI{8}{kpc}$ vs $>\SI{8}{kpc}$ & -1.122 & 0.250& 25\%\\
    \hline
    \end{tabular}
\end{table}

\endgroup

Comparing the distances obtained from these subsamples with the measurement we performed using the full sample shows that excluding the central PNe changes the final measurement only marginally. 
This is somewhat expected, since the distance measurement is mostly driven by the brightest nebulae and, as is clear from Fig.~\ref{fig:spatial}, they are not located in the centre of the galaxy.

The causes of the spatial variation of the PNLF could be several. 
If we exclude calibration issues, which are unlikely given the comparisons performed in Sec.~\ref{sec:rekola} and Sec.~\ref{sec:jacoby}, other possibilities could be internal extinction due to dust or variations in the gas-phase metallicity, with the first being the most likely explanation. 
These properties could also affect the discrepancy between the PNLF and TRGB distances, so we will analyse them in the following sections.

\subsection{Metallicity}
\label{sec:metallicity}

As introduced in Sec.~\ref{sec:pnlf}, empirical studies suggest that the zero point of the PNLF depends on the metallicity of the gas \citep[e.g.,][]{Dopita92,Ciardullo02,Scheuermann22}, and, in particular, it increases when the metalliticy decreases.
Given the shallow relation and the metallicity of \ngc253, very close to solar \citep[$12+\log(\rm O/H) = 8.69$][]{Beck22}, in Sec.~\ref{sec:pnlf} we ignored this effect, but here we will revise this assumption, considering also potential effects from the radial metallicity gradient of the galaxy. 

Considering a solar metallicity from \citet{Asplund09}, the change in zero-point vs metallicity is described by the following relation \citep{Ciardullo02, Scheuermann22}:
\begin{equation}
\label{eq:mstar}
    \Delta M^*_{5007} = 0.928[\mathrm{O/H}]^2 - 0.109[\mathrm{O/H}]+0.004.
\end{equation}
We can invert this equation to return the expected metallicity of the gas given the $\Delta M^*_{5007}$ needed to obtain a distance of \SI{3.5}{Mpc} from our PNLF.
We consider only the lower solution of the equation, given that the relation between $\Delta M^*_{5007}$ and metallicity has been characterised for metallicities lower than solar.
Obtaining a distance of \SI{3.5}{Mpc} from our PNe sample would require $M^*_{5007}=-4.23~\si{mag}$, which corresponds to $\Delta M^*_{5007} = 0.31$ and a metallicity of $12+\log(\rm O/H) = 8.17$ ($\sim 1/3~\si{\Zsun}$).
From the analysis of $\sim 1000$ \hii\ regions identified by McClain et al. (in prep) we recover a preliminary gradient of $-0.28~\si{dex/R_{25}}$ (Congiu et al., in prep.), which makes it impossible to reach an average metallicity of $12+\log(\rm O/H) = 8.17$ starting from the \citet{Beck22} nuclear value.
This gradient also rules out metallicity as the origin of the fainter PNLF observed in the central part of the galaxy, as it would require a heavily inverted gradient, which is clearly not observed.

\subsection{Extinction}
\label{sec:extinction}

\begin{figure*}
    \centering
    \includegraphics[width=\textwidth]{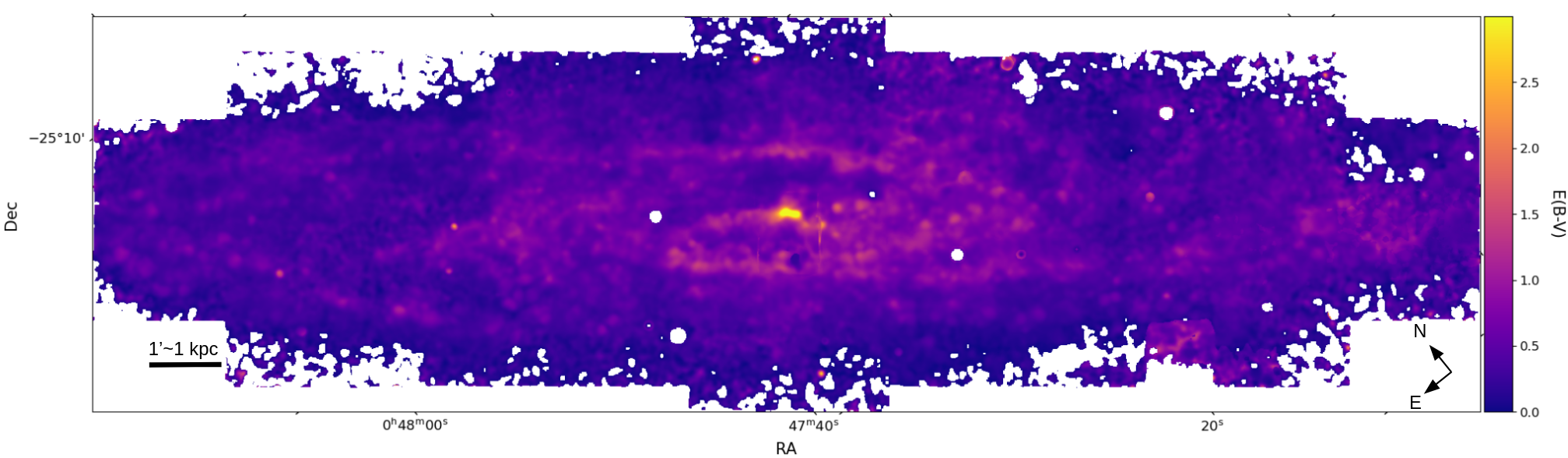}
    \caption{\ebv\ map of \ngc253. The \ebv\ has been calculated from the Balmer decrement using a convolved (to a \SI{5}{arcsec} FWHM Gaussian PSF) and binned version of the data, in order to detect both \Ha\ and \Hb\ across the majority of the FOV.}
    \label{fig:ebv}
\end{figure*}

When studying the PNLF of a galaxy, extinction is a delicate matter.
First, the foreground extinction caused by the Milky Way and the extinction caused by dust present in the host galaxy should be removed from the \oiii\ fluxes.
Such dust causes the PNe to appear fainter than they are, and in the most extreme cases it can also prevent the detection of bright objects located inside or behind the screen.
Second, intrinsic extinction, that is, the extinction caused by dust produced by the AGB star during the final stages of its evolution and located within the PN itself, has traditionally not been removed \citep{Ciardullo99,GarciaRojas18, Davis18}. 
As a result, the empirical functional form of the PNLF implicitly contains this bias, and any attempt to correct for intrinsic extinction may require a redetermination of the empirical functional form itself.

Removing the foreground Milky Way extinction is straightforward, as we describe in Sec.~\ref{sec:phot}.
However, correcting the internal extinction without removing the PNe intrinsic extinction is extremely challenging.
A standard approach such as estimating the \ebv\ of each nebula through the Balmer decrement of their spectrum is not an option, since it does not differentiate between the host galaxy extinction and the intrinsic PN extinction. 
In addition, simple arguments based on the vertical distribution of PNe and dust in galaxies suggest that, in most cases, the host extinction should not affect the PNLF \citep[e.g.,][]{Feldmeier97}.
Under typical conditions, there should be a large enough number of bright and unextinguished PNe to not affect the distance estimate, so only the foreground Galactic extinction is typically applied to the data.
However, \ngc253 is a highly inclined galaxy with prominent dust features clearly visible throughout the disk, so extinction could play a more significant role.
 
\subsubsection{Extinction Map}

In Fig.~\ref{fig:ebv} we show the \ebv\ map of \ngc253 recovered from the Balmer decrement.
To obtain it, we convolved our mosaic to an angular resolution of \SI{5}{\arcsec}\footnote{We assumed an average PSF FWHM of 0.8 across the mosaic (see Sec.~\ref{sec:psf}) and we convolved it using a 2D Gaussian Kernel with the FWHM needed to recover a final FWHM of \SI{5}{arcsec}. The spatial and spectral variation of the final FWHM is of the order of \SI{0.1}{\arcsec}} and binned it to \SI{0.8}{\arcsec\per\pix}, to ensure detection of \Ha\ and \Hb\ across most of the FOV.
We then estimated the \ebv\ assuming a theoretical \Ha$/$\Hb\ $=2.86$ \citep[Case B recombination][]{Osterbrock06}, a \citet{Cardelli89} extinction law and $\rm R_{V}=3.1$.
The resulting map shows a median \ebv\ of $\sim$ \SI{0.36}{mag} ($A(V) \sim$ \SI{1.1}{mag}), with peaks of $> $\SI{6}{mag} ($A(V) > $\SI{18}{mag}) close to the centre of \ngc253.
\citet{Rekola05} determined that for an extinction similar to our average one, the effect on the measured distance modulus should be of the order of a few \SI{0.1}{mag}\footnote{\citet{Rekola05} do not provide a model that we could test. Our analysis is based mainly on the discussion in their Sec.~3 and Fig. 4}.
The discrepancy between our distance modulus and that recovered by TRGB measurements from, for example, \citet{Okamoto24}, is $\sim$ \SI{0.35}{mag}, which is, indeed, in line with the prediction of \citet{Rekola05}.

\begin{figure}
    \centering
    \includegraphics[width=\linewidth]{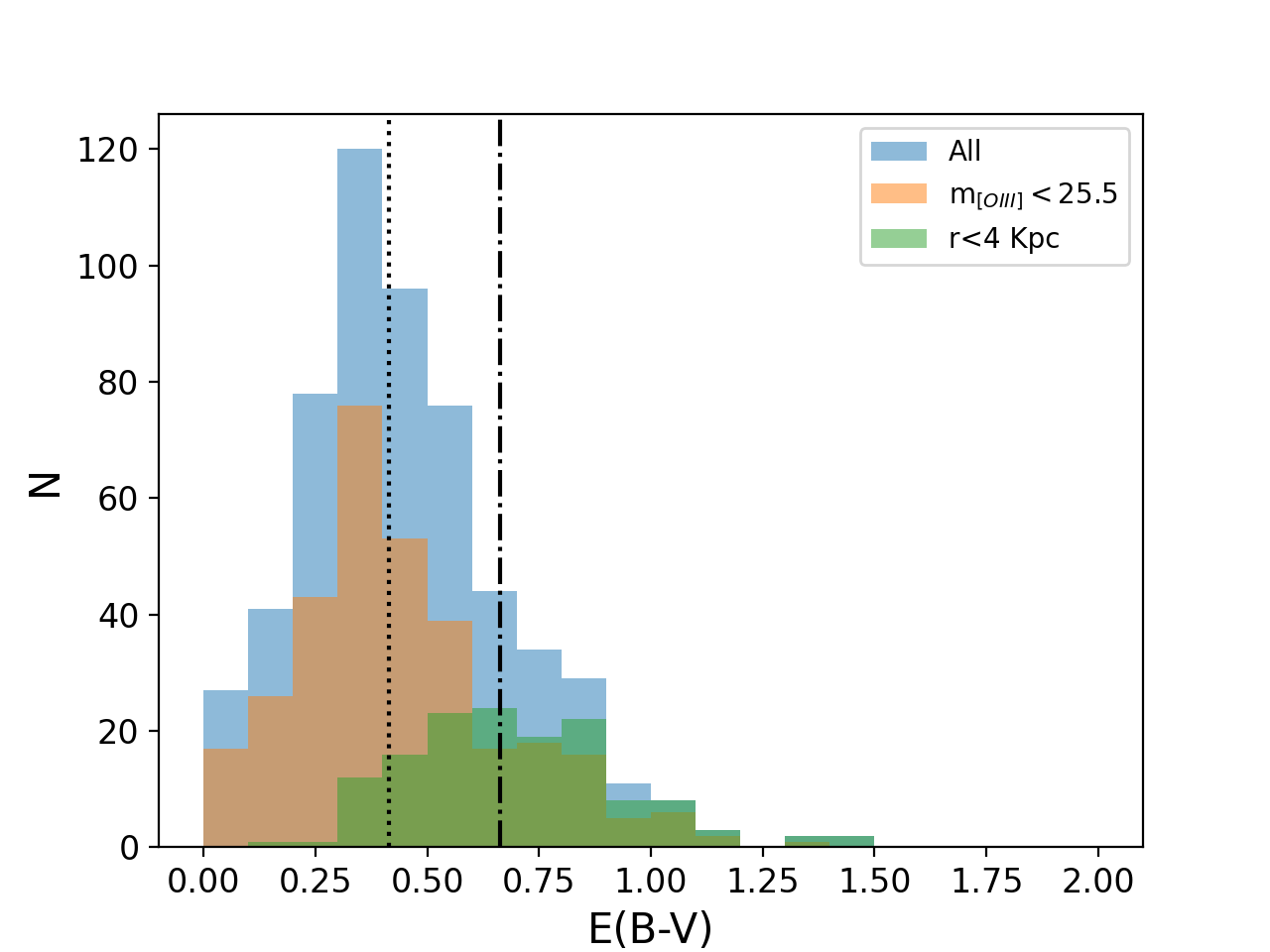}
    \caption{Maximum host galaxy \ebv\ associated with each PN. In blue we show the full sample, in orange only the PNe brighter than \SI{25.5}{mag}, and in green the PNe located within \SI{4}{kpc} from the centre of the galaxy. The dotted line represents the median \ebv\ for the full sample, while the dashed-dotted line shows the median for PNe in the centre of the galaxy.}
    \label{fig:ebv_hist}
\end{figure}

The MUSE extinction map represents the extinction produced by the dust screen in front of the \Ha\ and \Hb\ emitting gas.
In the case of a dusty interstellar medium with embedded emission, this represents approximately half of the total possible extinction.
We do not know the position of each PN along the line of sight, and we also know that PNe and dust typically follow different distributions, especially in the vertical direction, so we cannot use this map to directly correct the fluxes of our PNe.
Nevertheless, comparing the extinction map with the position of our PNe can still help us gauge its effect on the PNLF.
For each PN, we used the \SI{5}{arcsec} resolution extinction map shown in Fig.~\ref{fig:ebv} to recover the average \ebv\ contained in the same aperture that we used to correct the \oiii\ fluxes for background emission.
We do not simply measure the extinction at the exact location of our PNe to avoid including any effect of the intrinsic extinction, although this is probably not necessary given the resolution of our convolved data.
Figure~\ref{fig:ebv_hist} shows the distribution of \ebv\ associated with the PNe in the sample (blue histogram). 
We have a distribution ranging from \ebv~$=0$ to \ebv~$=$~\SI{1.44}{mag} ($A(V) \sim$ \SI{4.45}{mag}) with an average of \SI{0.45}{mag} and a median of \SI{0.41}{mag} ($A(V) \sim$ \SI{1.40}{mag} and \SI{1.27}{mag}, respectively).
This confirms that we are in a regime in which the effects of extinction on the PNLF are no longer negligible.
If we consider only the PNe brighter than \SI{25.5}{mag} (those included in the fit, orange histogram), we see that they have a very similar distribution.
Finally, in green, we show the \ebv\ distribution for PNe located within \SI{4}{kpc} of the nucleus of the galaxy.
As expected, this sample includes the objects in the high-extinction tail of the distribution, which could explain the significant difference between the distance estimate we obtain using only this sub-sample of sources (see Sec.~\ref{sec:radial}).

\subsubsection{Completeness}

\begin{figure}
    \centering
    \includegraphics[width=\linewidth]{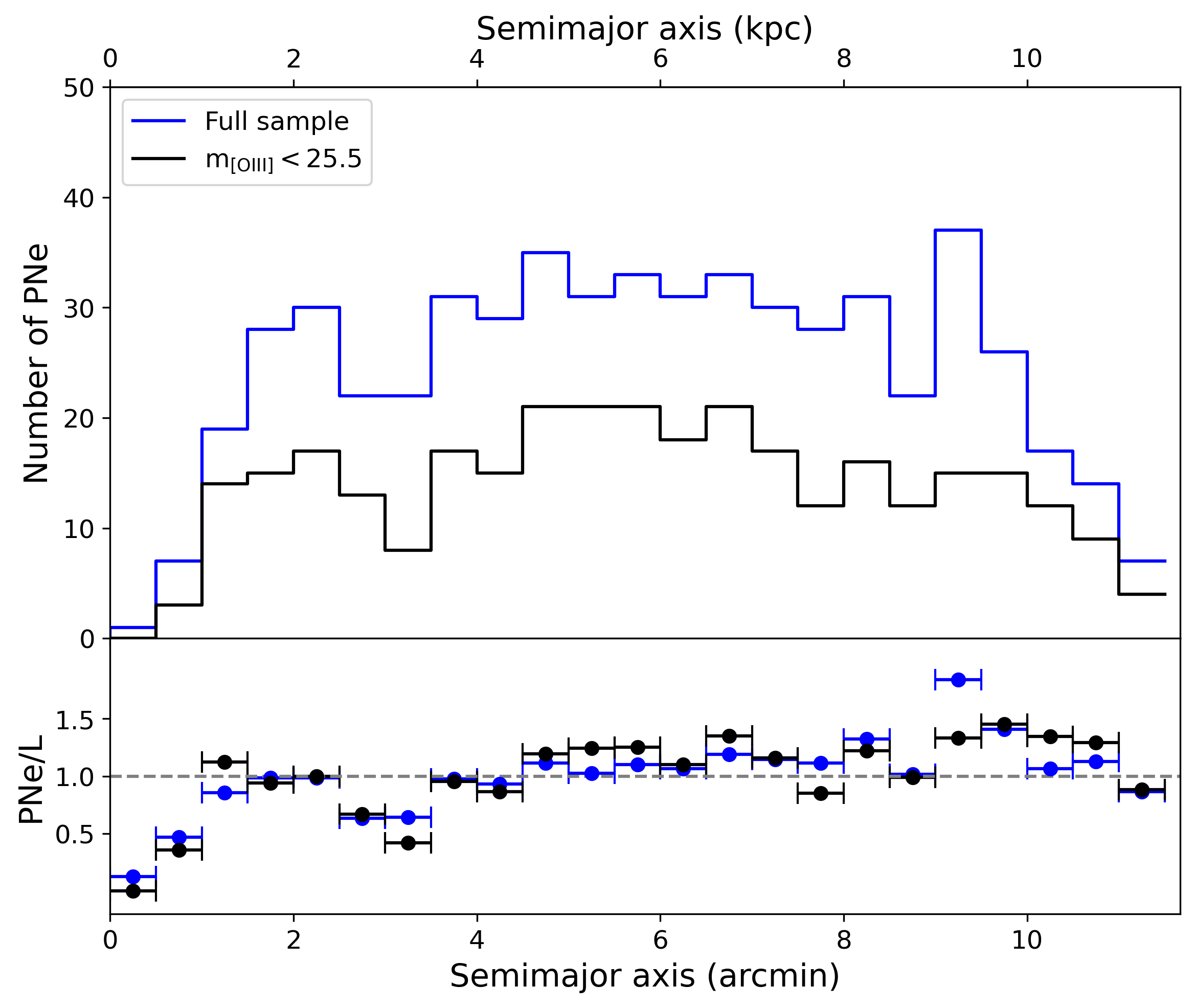}
    \caption{Distribution of PNe as a function of distance from the centre of the galaxy. The top panel shows the number of PNe contained in concentric elliptical rings with growing semimajor axis. The bottom panel shows the luminosity-specific PNe number density measured in the same rings and normalized of its average value, since we are interested in the trend and not on its absolute value. We show the full sample in blue and the sample with m$_{\rm [OIII]} <$~\SI{25.5}{mag} in black.}
    \label{fig:pne_lum}
\end{figure}

Another way the extinction could impact the PNLF is by affecting the completeness of the PNe sample, particularly in the dusty galaxy centre.
Figure~\ref{fig:pne_lum} shows how the number of PNe in concentric elliptical annuli changes as a function of distance from the centre of the galaxy both as an absolute measurement (top panel) and as the ratio between the number of PNe and the flux contained in the ellipse (bottom panel). 
To measure these quantities, we defined a series of ellipses centred on the centre of the galaxy and with constant PA and ellipticity and we counted how many PNe fell in each elliptical ring.
We then used the WFI R-band image presented in Sec.~\ref{sec:alignment} to extract the integrated flux in the same areas, masking bright stars, and we computed the the ratio between the number of PNe and the flux contained in each annulus\footnote{Since we are interested in the relative distribution of these points, and not in their absolute value, we did not convert the fluxes in luminosities.}.
This quantity has been shown to be quite independent from the properties of the stellar population of a galaxy \citep{Ciardullo89,Buzzoni06}, therefore any variations could indicate changes in the PNe sample completeness, perhaps due to dust extinction.
Figure~\ref{fig:pne_lum} shows that this quantity is relatively stable across the disk. 
Only rings with more that 50\% coverage in our mosaic are included in the plot.
At distances $<$~\SI{4}{kpc} however, we see we see a significant change in PNe/L, indicating a drop in completeness.
This happens both at the very centre, where we have the starburst ring and higher measured values of \ebv, and around $\sim$\SI{3.5}{kpc} from the centre, which correspond approximately to the end of the bar.
Here, both Fig.~\ref{fig:ngc253_vri} and Fig.~\ref{fig:ebv} show strong dust lanes and high extinction values.

\subsubsection{Modelling the effects of extinction on the PNLF}
\begin{figure}
    \centering
    \includegraphics[width=\linewidth]{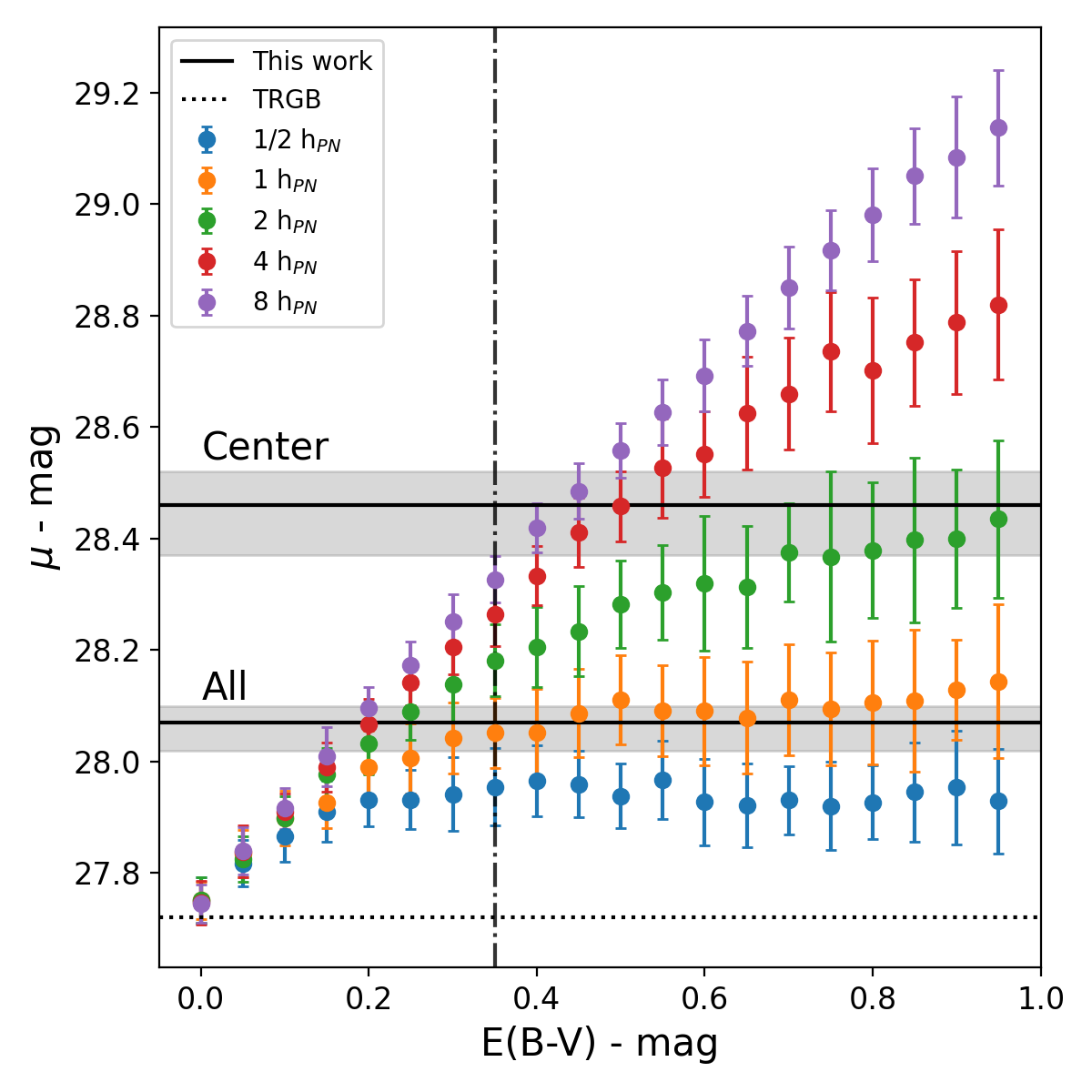}
    \caption{Effects of the extinction on the distance modulus measured via the PNLF. The points show how the distance modulus estimated via a sample of PNe extracted from a PNLF with fixed zero-point and distance modulus ($M^* = -4.54~\si{mag}$ and \SI{27.72}{mag} respectively) changes when applying an increasing amount of extinction. In this model, both PNe and extinction (dust) follow a Laplace distribution, as described in Sec.~\ref{sec:extinction}. Different colours represent different ratios between the PNe and dust scale height. The solid horizontal lines represent our measurement of the distance modulus of the galaxy when considering both the full sample (bottom line) and only the PNe close to the centre of the galaxy (top line), while the grey areas represent their uncertainties. The dashed horizontal line represent a distance modulus of \SI{27.72}{mag}, the expected one from the TRGB-based distances from the literature. Finally, the vertical dot-dashed line represents the average \ebv\ we measured from the extinction map shown in Fig.~\ref{fig:ebv}.}
    \label{fig:simple_models}
\end{figure}
In order to quantitatively assess the impact of dust extinction on the PNLF, we employ a straightforward model that describes the vertical distribution of dust and PNe within a disk to determine whether we could obtain a distance of \SI{4.10}{Mpc} by applying extinction to samples of PNe extracted from the expected PNLF of a galaxy situated \SI{3.5}{Mpc} away (\SI{27.72}{mag}), as is the case of \ngc253. 
Following the assumptions of \citet{Rekola05}, we consider that both PNe and dust are distributed following a Laplace distribution (also known as a double exponential distribution).
This represents their vertical distribution across the galactic disk, where the maximum of the probability coincides with the disk plane.
To keep the model simple, we assume a face-on disk, ignoring potential inclination effects.
We first extract a sample of PNe from the PNLF and assign them a location within the disk by extracting it from their Laplace distribution.
We then recover the expected extinction at their location from the cumulative distribution of dust normalised to the maximum extinction we considered. 
The extinction is then applied assuming a \citet{Cardelli89} extinction law, and an $\rm R_{V}=3.1$,
Finally, the modified PNLF was fitted with the method from Sec.~\ref{sec:pnlf}. 

We varied the maximum extinction from 0 to \ebv~$ = $~\SI{1}{mag}.
Then, we set a fixed scale height for the PNe and examined multiple scale heights for the dust: half, one, two, four, and eight times the PNe scale height. 
For each combination of parameters, we repeated the entire process 50 times, considering the mean distance as our measurement and its standard deviation as our error. 
The results of this analysis are presented in Fig.~\ref{fig:simple_models}. 
As expected, the extinction influences the distance inferred from the PNLF. 
Larger maximum extinctions result in more pronounced effects, making the galaxy appear further away, until a plateau is reached. 
Beyond this plateau, further increases in extinction do not affect the distance. 
The \ebv\ at which the plateau emerges depends on the relative scale heights of dust and PNe. 
When both are similarly distributed or when the dust has a higher scale height, we can recover our findings based on the anticipated PNLF of \ngc253. 
The \ebv\ at which this alignment occurs is approximately \SI{0.3}{mag} if dust and PNe share the same scale height, consistent with \citet{Rekola05}. 
If the dust scale height exceeds that of the PNe, we need only and \ebv\ of \SI{0.2}{mag} to explain our discrepancy.
Finally, if the PNe scale height is larger than the dust one, the plateau is reached at very low \ebv.
In this case, the difference between the measured distance modulus and the expected one is small and cannot explain our discrepancy. 
This result is in line with the argument exposed by \citet{Feldmeier97}, which states that in this scenario there should always be a large enough number of bright and unextincted PNe to not affect the final distance.

Extinction can also explain the difference between the distances recovered from disk and the central PNLFs.
In this case, Fig.~\ref{fig:simple_models} suggests that both a larger amount of extinction and a larger dust scale height are needed to explain our observations.
In particular, it shows that the dust needs to be distributed with a scale height that is at least twice as large as the PNe one, while the maximum extinction needs to be $\gtrsim 0.4$, even though it changes significantly depending on the dust scale height.

This analysis suggests that internal extinction within the host galaxy is responsible for the discrepancies we observe, as well as for the differences in zero-point between the central and the disk PNLF. 
In particular, it suggests that in \ngc253 the dust is distributed with a similar scale height relative to the PNe in the disk, and a much larger scale height if we consider only the region close to the centre of the galaxy.
This was not expected, since studies of both the Milky Way \citep{Allen73, Li18} and other edge-on nearby galaxies \citep[e.g.][]{Xilouris99, Bianchi07, DeGeyter14} show how they exhibit a thin disk of dust, with a scale height half that of the stellar disk (where the PNe reside).
In \ngc253 the starburst-driven outflow might be ejecting large quantities of dust-rich material from the centre of the galaxy, boosting the dust scale height.
This effect is stronger close to the outflow, which is why we need a much larger scale height to explain the distance recovered from the central PNLF.
The dust then settles when it moves further away from the outflow, but into a thicker disk with respect to more quiescent galaxies.

The proximity of \ngc253 and availability of highly precise distance measurements from multiple techniques show how the assumption that the PNLF is independent of the amount of dust in a galaxy is not always correct. 
As stated by \citet{Jacoby24}, having large and heterogeneous galaxy samples with well-sampled PNLFs is essential to characterise the circumstances where the dust distribution significantly biases the PNLF distance.
While correcting for the host galaxy dust extinction at any given position in the disk where a PN is found is challenging, our growing multi-wavelength views of the dust-rich ISM, along with the potential to develop rich kinematic models from the combined stellar and gas disk components, could be used to develop a statistical 3D approach to forward-model the line-of-sight dust extinction of the PNe population. 
Such an analysis is beyond the scope of this work but may present a novel way to refine the use of the PNLF as a rung in the cosmological distance ladder.

\section{Summary and Conclusions}
\label{sec:summary}

This work presents a new MUSE mosaic of the nearby starburst galaxy \ngc253.
The mosaic includes 103 MUSE pointings covering an approximate area of 20$\times$\SI{5}{armin^2} with $\sim$9 million spectra, and represents the largest mosaic of an extragalactic source ever observed by MUSE so far, not counting the sparse mosaic of the Large Magellanic Cloud \citep{Boyce17}.
The data were observed as part of two ESO programmes (108.2289 and 0102.B-0078(A)) for a total of $\sim53.5$ hrs of observing time.
Exposures were reduced using the official ESO MUSE pipeline \citep{Weilbacher20} through the \verb!pymusepipe! wrapper \citep{Emsellem22} and processed with the PHANGS data analysis pipeline, following the procedures described in \citet{Emsellem22}.
Special care has been given to the sky subtraction, given the low redshift of the galaxy.

We then exploit moment-zero maps of the main emission lines (\oiii, \Ha, and \sii) to identify a sample of 571 PNe.
We use 320 of them (those brighter than \SI{25.5}{mag}, our estimated completeness limit) to build the PNLF of the galaxy and recover its distance.
Following the procedure described in \citet{Scheuermann22}, we estimate a distance modulus of $28.07^{+0.03}_{-0.05}$~mag ($4.10^{+0.07}_{-0.09}$~Mpc), which is $\sim\SI{0.35}{mag}$ ($\SI{0.6}{Mpc}$) larger than expected from previous independent distance estimates from PNLF fitting and the TRGB method.
In particular, the TRGB-based distance ($\sim$\SI{27.72}{mag}, \SI{3.5}{Mpc}) has been independently confirmed several times over the last 15 years, proving itself to be a reliable reference.
The PNLF analysis revealed the following:
\begin{itemize}
    \item The distance recovered by \citet{Rekola05} ($27.62^{+0.16}_{-0.26}$~mag or $3.34^{+0.26}_{-0.38}$~Mpc) was driven by the misclassification of their brightest region, which now appears as an \hii\ regions. 
    Removing it from their sample, reconciles their distance with ours.
    \item The zero point of the relation changes significantly when considering only PNe located in the centre of the galaxy versus those located in the rest of the disk.
    In particular, the inner PNLF returns a distance that is $\sim$ \SI{0.45}{mag} (\SI{0.9}{Mpc}) larger than the distance recovered from the rest of the disk.
    \item We reject metallicity as the origin of the discrepancy that we observe between our measurement and the literature.
    Preliminary measurement of the gas phase metallicity and its gradient shows values corresponding to a negligible correction of the PNLF zero-point.
    \item We reject metallicity also as the origin of the spatial variation of the PNLF, since the observed behaviour would require an inverted metallicity gradient not observed in the galaxy.  
    \item A Balmer-decrement-based extinction map shows that the galaxy is affected by a significant amount of extinction.
    We estimate a median \ebv\ of $\sim \SI{0.36}{mag}$ ($A(V) \sim \SI{1.1}{mag}$), with peaks of $> \SI{6}{mag}$ ($A(V) > \SI{18}{mag}$) close to the \ngc253 centre.
    \item A simple model for the distribution of the PNe and dust shows that we can reproduce our results for the whole galaxy if we assume a maximum \ebv$\sim$\SI{0.3}{mag} and the same scale height for dust and PNe.
    Reproducing the estimate from the central PNLF, requires the dust to be distributed with at least twice the PNe scale height.
    This is in contrast with what is expected from scale height studies of the Milky Way and other nearby edge on galaxies \citep[e.g.][]{Allen73, Xilouris99, Bianchi07, DeGeyter14}, but it is compatible with a scenario where the starburst-driven outflow is enhancing the vertical distribution of dust in the \ngc253.
   
\end{itemize}
In conclusion, this work shows that the PNLF is not a reliable method to estimate the distance of \ngc253, because of its high dust content and its peculiar distribution.
In general, it shows that dust should not be ignored in the PNLF analysis, especially when working on galaxies characterised by high inclination and dust content, and that we need more galaxies with a well sampled PNLF to better characterise these effects.

\section*{Data availability}
Datacubes and DAP products are available at the CADC (\url{https://www.canfar.net/storage/vault/list/phangs/RELEASES/PHANGS-MUSE-NGC253}) and ESO Phase 3 archive (TBD).
The PNe catalogue is available only in electronic form at the CADC (\url{https://www.canfar.net/storage/vault/list/phangs/RELEASES/PHANGS-MUSE-NGC253/PNe_catalog}) or at the CDS via anonymous ftp to cdsarc.u-strasbg.fr (130.79.128.5) or via \url{http://cdsweb.u-strasbg.fr/cgi-bin/qcat?J/A+A/}.

\begin{acknowledgements}
The authors would like to thank our referee, Dr. Micheal Richer, for the excellent comments that helped improve this work. 
Based on observations collected at the European Organisation for Astronomical Research in the Southern Hemisphere under ESO programmes 108.2289 (PI: Congiu) and 0102.B-0078(A) (PI: Zschaechner).
This work was carried out as part of the PHANGS collaboration.
FS and KK acknowledge support from the Deutsche Forschungsgemeinschaft (DFG, German Research Foundation) in the form of an Emmy Noether Research Group (grant number KR4598/2-1, PI Kreckel) and the European Research Council’s starting grant ERC StG-101077573 (“ISM-METALS"). 
OE acknowledges funding from the Deutsche Forschungsgemeinschaft (DFG, German Research Foundation) -- project-ID 541068876.
E.C. and J.H. acknowledge the financial support from the visitor and mobility program of the Finnish Centre for Astronomy with ESO (FINCA).
JEMD acknowledges support from project UNAM DGAPA-PAPIIT IG 101025, Mexico.
A.D.B.~and S.A.C.~acknowledge support from the NSF under award AST-2108140.
R.S.K.\ acknowledges financial support from the ERC via Synergy Grant ``ECOGAL'' (project ID 855130),  from the German Excellence Strategy via the Heidelberg Cluster ``STRUCTURES'' (EXC 2181 - 390900948), and from the German Ministry for Economic Affairs and Climate Action in project ``MAINN'' (funding ID 50OO2206).  
RSK also thanks the 2024/25 Class of Radcliffe Fellows for highly interesting and stimulating discussions. 
This research has made use of the NASA/IPAC Extragalactic Database, which is funded by the National Aeronautics and Space Administration and operated by the California Institute of Technology.
This research has made use of the Astrophysics Data System, funded by NASA under Cooperative Agreement 80NSSC21M00561.
This research made use of Astropy, a community-developed core Python package for Astronomy \citep{Astropy13, Astropy18}, Matplotlib \citep{Hunter07}, MPDAF \citep{MPDAF16, MPDAF17}, NumPy \citep{Harris20}, Photutils \citep{Photutils24}, Pyneb \citep{Luridiana15}, SciPy \citep{Virtanen20}, Scikit-learn \citep{Scikit11}.
\end{acknowledgements}

\bibliographystyle{aa} 
\bibliography{bibliography}

\begin{appendix}

\section{DAOFIND based PNLF}
\label{app:daofind}
\FloatBarrier

\begin{figure}[h]
    \centering
    \includegraphics[width=\linewidth]{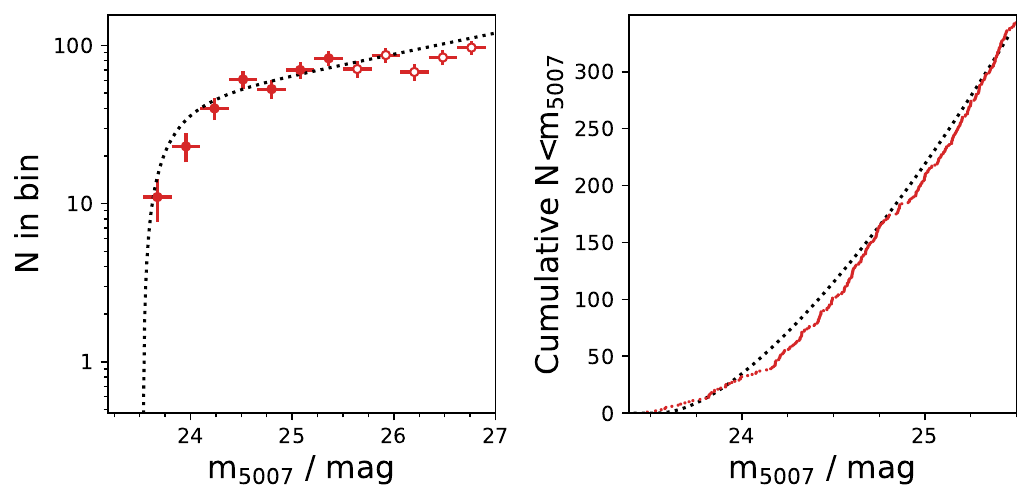}
    \caption{PNLF and cumulative luminosity function for \ngc253 using the sample of PNe recovered by DAOFIND. Panels, markers and colours as in Fig.~\ref{fig:pnlf}.}
    \label{fig:pnlf_daop}
\end{figure}

In this section we repeat the selection of the PNe candidates using the well-known point source detection algorithm DAOFIND \citep{Stetson87}.
We apply the algorithm in the version included in the \verb!photutils! package on the \oiii\ emission line map.
We used default arguments for the detection and assume a PSF size of $\SI{1}{arcsec}$.
This detection approach returns a catalogue of 10635 PNe candidates.
We then processed the catalogue of PNe candidates following the procedures described in Sec.~\ref{sec:analysis} to measure the FHWM of each source, the line fluxes, and to remove contaminants from the sample.
The final number of confirmed PNe is 5494, a rather high number compared to what identified in the emission line maps with the detection method described in Sec.~\ref{sec:detection}.
In fact, a visual inspection of the new candidates reveals that the vast majority of them are probably false detections.
However, if we consider only sources with $m_{\mathrm[OIII]} \leq 25.5~\si{mag}$, the number decreases to 343, quite comparable to the 319 PNe that result from visual inspection.
Matching the two catalogues, we see that 309 of the 320 confirmed PNe brighter than $\SI{25.5}{mag}$ are included in the clean sample of bright PNe from DAOFIND.

We used this clean sample of bright PNe to fit the PNLF (see Fig.~\ref{fig:pnlf_daop}).
As expected, given the similarity between the two samples, we obtain very similar results.
In particular, we obtain a distance modulus of $28.07^{+0.03}_{-0.4}$ mag, which corresponds to a distance of $4.12^{+0.07}_{-0.08}$ Mpc.
Although the final results of the two detection methods are similar, the large number of false detections that are classified as PNe when using the DAOFIND catalogue made us decide to rely on visual identification for the main analysis of the paper.

\FloatBarrier

\section{Additional tables.}
\label{app:add_tables}

\begingroup
\setlength{\tabcolsep}{10pt} 
\renewcommand{\arraystretch}{1.1} 
\begin{table*}[]
    \centering
    \small
    \caption{Summary of the observations.}
    \label{tab:summary_obs}
    \begin{tabular}{lrrrrrrr}
    \hline\hline
    OB Name & Date & Program & Airmass & Seeing & Exptime & FWHM1 & FWHM2\\
    \hline
    WFM-NGC-253-NW & 2019-07-29 & 0102.B-0078(A) & 1.06 & 0.80 & 1960.0 & 0.70& -\tablefootmark{a} \\
    WFM-NGC-253-SE & 2018-11-07 & 0102.B-0078(A) & 1.07 & 0.85 & 1960.0 & 0.72& -\tablefootmark{a} \\
    WFM-NGC253\_Pri01 & 2021-11-27 & 108.2289.001 & 1.21 & 0.58 & 844.8 & 0.74& 0.79\\
    WFM-NGC253\_Pri02 & 2021-11-30 & 108.2289.001 & 1.09 & 0.55 & 844.8 & 0.85& 0.82\\
    WFM-NGC253\_Pri03 & 2021-12-07 & 108.2289.001 & 1.32 & 0.84 & 844.8 & 0.99& 1.15\\
    WFM-NGC253\_Pri04 & 2021-12-28 & 108.2289.001 & 1.23 & 0.51 & 844.8 & 0.68& 0.76\\
    WFM-NGC253\_Pri05 & 2021-12-29 & 108.2289.001 & 1.22 & 0.77 & 844.8 & 0.91& 0.78\\
    WFM-NGC253\_Pri06 & 2022-06-29 & 108.2289.001 & 1.17 & 0.49 & 844.8 & 0.69& 0.84\\
    WFM-NGC253\_Pri07 & 2022-07-02 & 108.2289.001 & 1.33 & 0.57 & 844.8 & -\tablefootmark{b}& 0.50\\
    WFM-NGC253\_Pri08 & 2022-07-02 & 108.2289.001 & 1.13 & 0.53 & 844.8 & 0.63& 0.64\\
    WFM-NGC253\_Pri09 & 2022-07-04 & 108.2289.001 & 1.04 & 0.83 & 844.8 & 0.84& 0.72\\
    WFM-NGC253\_Pri10 & 2022-08-20 & 108.2289.001 & 1.04 & 0.55 & 844.8 & 0.71& 0.62\\
    WFM-NGC253\_Pri11 & 2022-08-27 & 108.2289.001 & 1.11 & 0.42 & 844.8 & 0.57& 0.60\\
    WFM-NGC253\_Pri12 & 2022-08-27 & 108.2289.001 & 1.02 & 0.85 & 844.8 & 0.70& 0.86\\
    WFM-NGC253\_Pri13 & 2022-08-27 & 108.2289.001 & 1.11 & 0.83 & 844.8 & 0.62& 0.69\\
    WFM-NGC253\_Pri14 & 2022-09-03 & 108.2289.001 & 1.04 & 0.82 & 844.8 & 0.76& 0.81\\
    WFM-NGC253\_Pri15 & 2022-09-03 & 108.2289.001 & 1.03 & 0.65 & 844.8 & 0.70& 0.60\\
    WFM-NGC253\_Pri16 & 2022-09-05 & 108.2289.001 & 1.08 & 0.73 & 844.8 & 0.64& 0.99\\
    WFM-NGC253\_Pri17 & 2022-09-05 & 108.2289.001 & 1.24 & 0.93 & 844.8 & 0.69& 0.89\\
    WFM-NGC253\_Pri18 & 2022-09-06 & 108.2289.001 & 1.22 & 0.74 & 844.8 & 0.67& 1.50\\
    WFM-NGC253\_Pri18 & 2023-01-16 & 108.2289.001 & 1.53 & 0.52 & 844.8 & 0.72& 0.74\\
    WFM-NGC253\_Pri19 & 2022-09-19 & 108.2289.001 & 1.03 & 1.04 & 844.8 & 1.32& 1.15\\
    WFM-NGC253\_Pri20 & 2022-09-19 & 108.2289.001 & 1.12 & 1.03 & 844.8 & 1.01& 0.90\\
    WFM-NGC253\_Pri21 & 2022-09-19 & 108.2289.001 & 1.30 & 1.00 & 844.8 & 0.95& -\tablefootmark{b}\\
    WFM-NGC253\_Pri22 & 2022-09-20 & 108.2289.001 & 1.03 & 1.07 & 844.8 & 1.25& 1.04\\
    WFM-NGC253\_Pri23 & 2022-09-20 & 108.2289.001 & 1.02 & 1.63 & 844.8 & 1.31& 1.61\\
    WFM-NGC253\_Pri24 & 2022-09-21 & 108.2289.001 & 1.10 & 0.60 & 844.8 & 0.67& 0.85\\
    WFM-NGC253\_Pri25 & 2022-09-22 & 108.2289.001 & 1.17 & 0.78 & 844.8 & 1.01& 0.96\\
    WFM-NGC253\_Pri26 & 2022-09-22 & 108.2289.001 & 1.09 & 0.74 & 844.8 & -\tablefootmark{b}& 0.96\\
    WFM-NGC253\_Pri27 & 2022-09-23 & 108.2289.001 & 1.33 & 0.98 & 844.8 & 0.84& 1.13\\
    WFM-NGC253\_Pri28 & 2022-09-23 & 108.2289.001 & 1.13 & 0.85 & 844.8 & 0.90& 1.16\\
    WFM-NGC253\_Pri29 & 2022-09-23 & 108.2289.001 & 1.03 & 0.94 & 844.8 & 1.11& 0.88\\
    WFM-NGC253\_Pri30 & 2022-09-23 & 108.2289.001 & 1.03 & 0.95 & 844.8 & 0.78& 0.73\\
    WFM-NGC253\_Pri31 & 2022-09-24 & 108.2289.001 & 1.23 & 1.24 & 844.8 & 1.18& 0.58\\
    WFM-NGC253\_Pri32 & 2022-09-24 & 108.2289.001 & 1.07 & 0.71 & 844.8 & 0.63& 0.66\\
    WFM-NGC253\_Pri33 & 2022-09-24 & 108.2289.001 & 1.21 & 0.60 & 844.8 & 0.56& 0.64\\
    WFM-NGC253\_Pri34 & 2022-09-25 & 108.2289.001 & 1.08 & 1.03 & 844.8 & 1.13& 0.82\\
    WFM-NGC253\_Pri35 & 2022-09-25 & 108.2289.001 & 1.03 & 0.58 & 844.8 & 0.51& 0.65\\
    WFM-NGC253\_Pri36 & 2022-09-26 & 108.2289.001 & 1.16 & 0.82 & 844.8 & 0.88& 0.84\\
    WFM-NGC253\_Pri37 & 2022-09-26 & 108.2289.001 & 1.04 & 0.98 & 844.8 & 0.98& 0.79\\
    WFM-NGC253\_Pri38 & 2022-09-26 & 108.2289.001 & 1.05 & 0.70 & 844.8 & 0.59& 0.76\\
    WFM-NGC253\_Pri39 & 2022-09-29 & 108.2289.001 & 1.03 & 0.73 & 844.8 & 0.64& 0.65\\
    WFM-NGC253\_Pri40 & 2022-09-30 & 108.2289.001 & 1.13 & 0.68 & 844.8 & 0.71& 0.72\\
    WFM-NGC253\_Pri41 & 2022-09-30 & 108.2289.001 & 1.03 & 0.80 & 844.8 & 0.61& 0.87\\
    WFM-NGC253\_Pri42 & 2022-09-30 & 108.2289.001 & 1.02 & 1.03 & 844.8 & 0.89& 0.83\\
    WFM-NGC253\_Pri43 & 2022-10-28 & 108.2289.001 & 1.05 & 0.70 & 844.8 & 0.68& 0.70\\
    WFM-NGC253\_Pri44 & 2022-08-26 & 108.2289.001 & 1.04 & 0.47 & 844.8 & 0.49& 0.58\\
    WFM-NGC253\_Pri45 & 2022-01-04 & 108.2289.001 & 1.34 & 0.74 & 844.8 & 0.84& 0.71\\
    WFM-NGC253\_Pri46 & 2022-10-28 & 108.2289.001 & 1.24 & 0.57 & 844.8 & 0.69& 0.76\\
    WFM-NGC253\_Pri47 & 2022-10-30 & 108.2289.001 & 1.03 & 0.57 & 844.8 & 0.81& 0.74\\
    WFM-NGC253\_Pri48 & 2022-10-30 & 108.2289.001 & 1.04 & 0.41 & 844.8 & 0.58& 0.60\\
    WFM-NGC253\_Pri49 & 2022-10-30 & 108.2289.001 & 1.14 & 0.52 & 844.8 & 0.66& 0.63\\
    WFM-NGC253\_Pri51 & 2023-05-30 & 108.2289.001 & 1.39 & 0.79 & 844.8 & 0.84& -\tablefootmark{a}\\
    \hline
    \end{tabular}
    \tablefoot{
    We report the name of the OB, the date of observation, the program ID, the average airmass of the observations, the average seeing estimated from the Paranal DIMM, the exposure time, and the average full width at half maximum (FWHM) of the point spread function (PSF) for each pointing observed in the OB as derived from the PNe in Sec.~\ref{sec:psf}. All OBs except for WFM-NGC-253-NW, WFM-NGC-253-SE, and WFM-NGC253\_Pri51, include two separate pointings (see Sec~\ref{sec:data}). The exposure time refers to the single pointing.
\tablefoottext{a}{These OBs include only a single pointing.}
\tablefoottext{b}{No PNe in the field of view of these pointings.}
}
\end{table*}
\endgroup

\label{sec:dptables}
\begin{table*}
\caption{Summary of the information included in the PNe catalogue.}
 \label{tab:cat_example}
\centering
\begin{tabular}{ll}
\hline\hline
\multicolumn{2}{c}{Position and Morphology}\\
\hline
\texttt{ID}& Identification number\\
\texttt{RA}& Right ascension of the nebula\\
\texttt{DEC}& Declination of the nebula\\
\texttt{deproj\_dist}& Deprojected distance from the nucleus of the galaxy (in arcsec)\\
\texttt{deproj\_phi}& Deprojected position angle (in degrees)\\
\texttt{FWHM}& FWHM of the nebula from the Moffat fit (in arcsec)\\
\hline\hline
\multicolumn{2}{c}{Spectral properties}\\
\multicolumn{2}{c}{\small \texttt{lineid} should be replaced with HEII4686, HB4861, HA6562, HEI5875\tablefootmark{a}, OIII5006, NII6583, SII}\tablefootmark{b}\\
\hline
\texttt{lineid\_FLUX}& Background subtracted line flux\tablefootmark{c} in $10^{-20}\,\si{erg.cm^{-2}.s^{-1}}$\\
\texttt{lineid\_FLUX\_ERR}& Line flux error in $10^{-20}\,\si{erg.cm^{-2}.s^{-1}}$\\
\texttt{lineid\_FLUX\_BACK}& Background flux in $10^{-20}\,\si{erg.cm^{-2}.s^{-1}}$\\
\texttt{lineid\_SNR}& \sn\ of the line\\
\texttt{m5007}&  $m_{5007}$ in magnitudes\\
\texttt{dm5007}&  Error on $m_{5007}$ in magnitudes\\

\hline\hline
\end{tabular}
\tablefoot{
\tablefoottext{a}{The \hei\ line is masked by the MUSE notch filter in the two central pointings observed with AO.}
\tablefoottext{b}{The \sii\ fluxes were recovered from a single moment map, given the vicinity of the two lines.}
\tablefoottext{c}{Line fluxes are corrected for Milky Way extinction and aperture size, as described in Sec.~\ref{sec:phot}}
}
\end{table*}

\begingroup

\begin{table*}
	
	\centering

    \caption{Wavelengths and ionisation potential of the relevant ion for each emission line.}
    \label{tab:ems}
	\begin{tabular}{lclcc}
		\hline
		Line name & Wavelength & String ID & Ionisation potential & Fixed ratio \\
                  & (air) [\AA] &          & [eV]                 & \\
		\hline
		\multicolumn{5}{c}{Hydrogen Balmer lines} \\
		\hline
		\Hb\ & 4861.35 & \texttt{HB4861} & 13.60 & no \\
		\Ha\ & 6562.79 & \texttt{HA6562} & 13.60  & no\\
		\hline
		\multicolumn{5}{c}{Low ionisation lines} \\
		\hline
        \niion$\lambda$5197 & 5197.90  & \texttt{NI5197}  & --- & no\\
		\niion$\lambda$5200 & 5200.26  & \texttt{NI5200}  & --- & no \\
		\niion$\lambda$5754 & 5754.59  & \texttt{NII5754} & 14.53 & no\\
		\niion$\lambda$6548 & 6548.05  & \texttt{NII6548} & 14.53 & 0.34 \nii\\
		\niion$\lambda$6584 & 6583.45  & \texttt{NII6583} & 14.53 & no \\
		\siion$\lambda$6717 & 6716.44  & \texttt{SII6716} & 10.36 & no \\
		\siion$\lambda$6731 & 6730.82  & \texttt{SII6730} & 10.36 & no\\
		\hline
		\multicolumn{5}{c}{High ionization lines} \\
		\hline
        \heiion$\lambda$4686  & 4685.70 & \texttt{HEII4865}  & 24.58 & no\\ 
		\oiiion$\lambda$4959 & 4958.91 & \texttt{OIII4958} & 35.12 & 0.35 \oiii \\ 
		\oiiion$\lambda$5007 & 5006.84 & \texttt{OIII5006} & 35.12 & no\\ 
		\heion$\lambda$5876 & 5875.61 & \texttt{HEI5875}  &  ---   & no\\
		\siiion$\lambda$6312 & 6312.06 & \texttt{SIII6312} & 23.34 & no\\
		\hline
	\end{tabular}
    \tablefoot{Wavelengths are taken from the National Institute of Standards and Technology (NIST; \url{https://physics.nist.gov/PhysRefData/ASD/lines_form.html}), and are Ritz wavelengths in air except for the H~Balmer lines, in which case we use the `observed ' wavelength in air as reported in NIST. The DAP string name is used to identify the correct extension in the MAPS files or in the moment maps. Ionisation potentials are taken from \cite{Draine11}. The \oion\ lines at 6300 and \SI{6363}{\angstrom} are not included in this list because they are heavily contaminated by sky emission.}
\end{table*}

\renewcommand{\arraystretch}{1.1} 
\begin{table*}
\caption{List of \texttt{FITS} extensions included in the \texttt{MAPS} file}              
\label{tab:maps}      
\centering    
\begin{tabular}{l l}
\hline
\hline
Extension name &  Description\\
\hline
\hline
    \multicolumn{2}{c}{Binning} \\
\hline
	\texttt{FLUX} & white-light image\\
	\texttt{SNR} & continuum S/N ratio for individual spaxels \\
	\texttt{SNRBIN} & continuum S/N for each Voronoi bin \\
	\texttt{BIN\_ID} & unique ID for each Voronoi bin, unbinned spectra have bin IDs of $-1$ \\
\hline
	\multicolumn{2}{c}{Stellar kinematics} \\
	\multicolumn{2}{c}{\texttt{HN$^{\#}$\_STARS} = higher order Gauss-Hermite velocity moment, if available (e.g., \texttt{H3\_STARS, H4\_STARS})}\\
\hline
	\texttt{V\_STARS} & stellar velocity [$\rm km  \ s^{-1}$], after subtracting the systemic velocity\\
	\texttt{FORM\_ERR\_V\_STARS} & formal velocity error [$\rm km \ s^{-1}$]\\
	\texttt{SIGMA\_STARS} & stellar velocity dispersion [$\rm km \ s^{-1}$]\\
	\texttt{FORM\_ERR\_SIGMA\_STARS} & formal sigma error [$\rm km  \ s^{-1}$] \\
	\texttt{ERR\_SIGMA\_STARS} & MCMC-calculated error for sigma (if available) [$\rm km \ s^{-1}$]\\
	\texttt{HN$^{\#}$\_STARS} & higher order moments of the stellar LOSVD (when available) \\
	\texttt{FORM\_ERR\_HN$^{\#}$\_STARS}  & formal errors in the high-order moments \\
	\texttt{ERR\_HN$^{\#}$\_STARS} & MCMC errors for higher order moments (not yet available)  \\
\hline
	\multicolumn{2}{c}{Stellar populations} \\
\hline
	\texttt{STELLAR\_MASS\_DENSITY}  &  stellar mass surface density [$\rm M_\odot \ pc^{-2}$] \\
	\texttt{STELLAR\_MASS\_DENSITY\_ERR}  &  error in the above [$\rm M_\odot \ pc^{-2}$] \\
	\texttt{AGE\_MW}  & log(Age/yr), where the Age is mass-weighted \\
	\texttt{AGE\_MW\_ERR}  & error in the above quantity\\
	\texttt{Z\_MW}  &  mass-weighted [Z/H] \\
	\texttt{Z\_MW\_ERR}  &  error in the above quantity \\
	\texttt{AGE\_LW}  &  log(Age/yr), where the Age is luminosity-weighted (V-band)  \\
	\texttt{AGE\_LW\_ERR}  &  error in the above\\
	\texttt{Z\_LW}  & luminosity-weighted (V-band) [Z/H]  \\
	\texttt{Z\_LW\_ERR}  & error in the above quantity  \\
	\texttt{EBV\_STARS}  & $E(B-V)$ of the stellar component [mag] \\
\hline
	\multicolumn{2}{c}{Emission lines} \\
	\multicolumn{2}{c}{ \texttt{*emline} = emission line string id listed in Table~\ref{tab:ems} }\\
\hline
	\texttt{BIN\_ID\_LINES}  & unique bin for emission lines, these are individual spaxels in the current DR2 
	\\
	\texttt{CHI2\_TOT}  & The $\chi^2$ over the full fitted wavelength range. 
	\\
	\texttt{*emline\_FLUX}  & emission line flux [$\rm 10^{-20} \ erg \ s^{-1} \ cm^{-2}  \ spaxel^{-1}$] \\
	\texttt{*emline\_FLUX\_ERR}  & emission line flux error [$\rm 10^{-20} \ erg s^{-1} cm^{-2} \ spaxel^{-1}$] \\
	\texttt{*emline\_VEL}  & emission line velocity [$\rm km \ s^{-1}$] \\
	\texttt{*emline\_VEL\_ERR} & emission line velocity error [$\rm km \ s^{-1}$]\\
	\texttt{*emline\_SIGMA} & emission line velocity dispersion [$\rm km \ s^{-1}$] \\
	\texttt{*emline\_SIGMA\_ERR} & emission line velocity dispersion error [$\rm km \ s^{-1}$] \\
	\texttt{*emline\_SIGMA\_CORR} & instrumental velocity dispersion at the position of the line [$\rm km \ s^{-1}$]\\
	\texttt{*emline\_MOM0}  & emission line moment 0 [$\rm 10^{-20} \ erg \ s^{-1} \ cm^{-2}  \ spaxel^{-1}$] \\
	\texttt{*emline\_MOM0\_ERR}  & emission line moment 0 error [$\rm 10^{-20} \ erg s^{-1} cm^{-2} \ spaxel^{-1}$] \\
	\texttt{*emline\_MOM1}  & emission moment 1 [$\rm km \ s^{-1}$] \\
	\texttt{*emline\_MOM1\_ERR} & emission moment 1 error [$\rm km \ s^{-1}$]\\
	\texttt{*emline\_MOM2} & emission moment 2 [$\rm km \ s^{-1}$] \\
	\texttt{*emline\_MOM2\_ERR} & emission moment 2 error [$\rm km \ s^{-1}$] \\    
\hline
\end{tabular}
\tablefoot{Each extension is a two-dimensional map on the same WCS as the mosaic datacube. We list the extension names, and a brief description of the map associated with that extension. All lines maps produced are corrected for the Milky Way foreground contribution. Moment-zero maps are not.}
\end{table*}
\endgroup

\end{appendix}

\end{document}